\documentclass[12pt]{article}

\usepackage{setspace}
\usepackage{amsmath,amsthm,amssymb,bbm} 
\usepackage{graphicx} 
\usepackage[round]{natbib} 
\usepackage{enumerate,color}
\usepackage{subfigure}
\usepackage{hyperref}

\def \tpr{\text{TPR}}
\def \ppv{\text{PPV}}
\def \ap{\text{AP}}
\def \auc {\text{AUC}}
\def \bs{\text{BrS}}
\def \sbs {\text{sBrS}}

\def \aphat{\widehat{\ap}}

\def \auchat{\widehat{\auc}}
\def \bshat{\widehat{\bs}}
\def\sbshat{\widehat{\sbs}}
\def \sumin{\sum_{i=1}^n}
\def \sumjn{\sum_{j=1}^n}

\def \bX {\bf X}

\def \bPsi {\boldsymbol{\Psi}}
\def \betahat{\widehat{\beta}}
\def \phat {\widehat{p}}
\def \rhat{\widehat{r}}
\def \pihat{\widehat{\pi}}
\def \fpr {\text{FPR}}
\def \tpr {\text{TPR}}
\def \ppv{\text{PPV}}
\def \Dfrak{\mathfrak{D}}
\def \sbX{\mbox{\tiny $\bX$}}
\newtheorem{remark}{Remark}

\title{Is the new model better? One metric says yes, but the other says no. Which metric do I use?}
\author{Qian M. Zhou$^{a,1,2}$, Zhe Lu$^{b}$, Russell J. Brooke$^{c}$,  Melissa M Hudson$^{c}$, Yan Yuan$^{b,2}$ \vspace{0.5cm} \\
\small $^{a}$ Department of Mathematics and Statistics, Mississippi State University,\\
\small  Starkville, Mississippi 39762, USA \\
\small $^{b}$ School of Public Health, University of Alberta, \\
\small Edmonton, AB T6G1C9, Canada\\
\small $^{c}$ St Jude Children’s Research
Hospital, Memphis, TN 38105, USA \\
}
\date{}
\begin{document}
\maketitle

\begin{abstract}
\noindent \textbf{Background:} Incremental value (IncV) evaluates the performance change from an existing risk model to a new model. It is one of the key considerations in deciding whether a new risk model performs better than the existing one. Problems arise when different IncV metrics contradict each other. For example, compared with a prescribed-dose model, an ovarian-dose model for predicting acute ovarian failure has a slightly lower area under the receiver operating characteristic curve (AUC) but increases the area under the precision-recall curve (AP) by 48\%. This phenomenon of conflicting conclusions is not uncommon, and it creates a dilemma in medical decision making. \medskip

\noindent \textbf{Methods:} In this article, we examine the analytical connections and differences between two IncV metrics: IncV in AUC ($\Delta\auc$) and IncV in AP ($\Delta\ap$). Additionally, since they are both semi-proper scoring rules, we compare them with a strictly proper scoring rule: the IncV of the scaled Brier score ($\Delta\sbs$), via a numerical study \medskip

\noindent \textbf{Results:} We demonstrate that both $\Delta \auc$ and $\Delta \ap$ are weighted averages of the changes (from the existing model to the new one) in separating the risk score distributions between events and non-events. However, $\Delta\ap$ assigns heavier weights to the changes in the high-risk group, whereas $\Delta\auc$ weights the changes equally. In the numerical study, we find that $\Delta\ap$ has a wide range, from negative to positive, but the size of $\Delta\auc$ is much smaller. In addition, $\Delta \ap$ and $\Delta\sbs$ are highly consistent, but $\Delta \auc$ is negatively correlated with $\Delta \sbs$ and $\Delta\ap$ at a low event rate. $\Delta\auc$ and $\Delta\ap$ are the least consistent among the three pairs, and their differences are more pronounced as the event rate decreases. \medskip

\noindent \textbf{Conclusions:} Choosing which metric to evaluate the IncV of a new model depends on the purpose of the prediction. If the risk model is used to identify the high-risk group, $\Delta \ap$ is more appropriate, especially for a low event rate.
\end{abstract}

\paragraph{Keywords:} High-risk group identification; Prediction performance; Brier score; Proper scoring rules; Rare outcome

\footnotetext[1]{Dr. Zhou is the corresponding author.}

\section{Introduction}

Risk prediction is crucial in many medical decision-making settings, such as managing disease prognosis. Numerous research has been dedicated to continually updating risk models for better prediction accuracy. For example, several papers have investigated the improvement in predicting the risk of cardiovascular disease by adding new biomarkers to the existing Framingham risk model, such as the C-reactive protein \citep{cook2006effect, buckley2009c}, and more recently, a polygenic risk score \citep{mosley2020predictive, elliott2020predictive}.

In some applications, an existing marker is replaced with a new marker that provides more precise information. For example, cancer  treatment such as radiation can have significant long-term health consequences for cancer survivors. Prescribed radiation doses to body regions, such as the abdomen and chest, are routinely available in medical charts.  But to predict the risk of an organ-specific outcome, e.g., secondary lung cancer or ovarian failure, a more precise measurement of the radiation exposure to specific organs provides better information. Radiation oncologists developed and applied algorithms to estimate these organ-specific exposures  \citep{howell2019adaptations}.

The measurement of a new marker or the more precise measurement of a known risk factor is often costly and time-consuming. Thus, it is important to verify that the new model indeed has a measurable and better prediction performance than the existing one, and thus, worth the extra resources. A number of metrics have been proposed to evaluate the incremental value (IncV) of the risk model that incorporates the new information. The IncV has primarily been discussed in settings where new markers are added to the existing risk profile \citep{pencina2008evaluating,pepe2013testing}. In this paper, the term IncV refers to the change of the prediction performance whenever an existing risk model is compared with a new one.

In medical research, the receiver operating characteristic (ROC) curve has been the most popular tool for model evaluation, dating back to the 1960s when it was applied in diagnostic radiology and imaging systems \citep{zweig1993receiver, pepe2003statistical}. The area under the ROC curve (AUC) captures the discriminatory ability of a model, i.e., how well a model separates events (subjects who experience the event of interest) from non-events (subjects who are event-free).  Recently, the precision-recall (PR) curve is gaining popularity \citep{badawi2018evaluation,chaudhury2019alzheimer,tang2019prediction,xiao2019comparison}. Originated from the information retrieval community in the 1980s \citep{raghavan1989critical,manning1999foundations}, it is a relatively new tool in medical research. The area under the PR curve is called the average positive predicted value or the average precision (AP) \citep{yuan2015threshold,su2015relationship,yuan2018threshold}. The AP evaluates the \textit{prospective} prediction performance of a risk model because it is conditional on the risk score obtained at baseline. In contrast, the AUC is a \textit{retrospective} metric that is conditional on future disease outcomes, unknown at baseline.

Problems arise when the IncV in AUC and IncV in AP contradict each other, which is not uncommon. For example, a prescribed-dose model and an ovarian-dose model were compared in Clark et al. (2020) for predicting acute ovarian failure among female childhood cancer survivors. The ovarian-dose has a slightly lower AUC but an increased AP by about 48\%, compared to the prescribed-dose model. Our numerical study results further demonstrate that the IncV in AUC and the IncV in AP do not agree in various scenarios when neither of the two competing working risk models is the true model. The disagreement creates confusion in decision making. In this article, we present explanations via unveiling their analytical relation and differences. 

\section{Notation, Definitions, and Data example}\label{sec:method}

First, we lay out the notations and define concepts that are used throughout this article. Let $D=$ 0 or 1 denote a binary outcome. For studies with an event time $T$, define $D=I(T \leq \tau)$ for a given prediction time period $\tau$, which indicates that the outcome is time-dependent. In this article, we refer to subjects with $D=1$ as the \textit{events} and those with $D=0$ as the \textit{non-events}. Let $\pi=Pr(D=1)$ denote the event rate. 

\subsection{Risk model and risk score}\label{sec:riskmodel}
A risk model is a function of a set of predictors $\bX=(X_1,\cdots,X_{k-1})$, which might include interaction terms and polynomial terms, to obtain the probability of $D=1$. Usually, we write this model as a regression model: 
\begin{equation}\label{equ:riskmodel}
p(\bX) = g(\beta_0+\beta_1X_1+\cdots+\beta_{k-1}X_{k-1}), 
\end{equation}
where $g(\cdot)$ is a smooth and monotonic \textit{link} function, such as a logit link. For the censored event time outcomes, a risk model could be Cox's proportional hazards model \citep{cox1972regression} or the time-specific generalized linear model \citep{uno2007evaluating}; both models can be expressed in the general form of equation (\ref{equ:riskmodel}) with modifications.

In practice, the underlying data generating mechanism is often complicated, and our working risk model in equation (\ref{equ:riskmodel}) is usually misspecified. Let $\pi(\bX) = Pr(D=1\mid \bX)$ denote the \textit{true} probability of $D=1$, which is determined by the underlying distribution of $D$ given $\bX$. Here, we refer to $\pi(\bX) $ as the \textit{true} risk and $p(\bX)$ as the \textit{working} risk from a working risk model. When the working risk model in equation (\ref{equ:riskmodel}) is misspecified, $p(\bX) \neq \pi(\bX)$. 

The working risk $p(\bX)$ can be regarded as a risk score and used to classify subjects into different risk categories. For example, given a cut-off value $c$, subjects with $p(\bX)\leq c$ are classified into the low-risk group, whereas the high-risk group consists of subjects with $p(\bX)>c$. In general, a risk score, denoted as $r(\bX)$, can be any function of $\bX$ that reflects how likely a subject is an event. Thus, $r(\bX)$ can be a non-decreasing transformation of $p(\bX)$, e.g., $r(\bX) = g^{-1}\left(p(\bX)\right) = \beta_0+\beta_1X_1+\cdots+\beta_{k-1}X_{k-1}$. 
\medskip

\begin{remark} \label{rem:true-risk}
In practice, the working risk $p(\bX)$ is estimated from a data sample. The estimated regression coefficients $\widehat \beta_j$, $j=0,1,\cdots,k-1$, are the solution to an estimating equation: $\bPsi(\beta_0,\cdots,\beta_{k-1}) = \sum_{i=1}^n \Psi(\beta_0,\cdots,\beta_{k-1};D_i,\bX_i)$.  The estimated risk given $\bX$ is  $\phat(\bX)=g(\betahat_0+\betahat_1X_1 + \cdots + \betahat_{k-1}X_{k-1})$, which is not of interest here. In this article, we investigate the predictive performance of the \textbf{population} working risk $p(\bX)=g(\beta_0^{\ast}+\beta_1^{\ast}X_1 + \cdots + \beta_{k-1}^{\ast}X_{k-1})$ where $\beta_j^{\ast}$'s are the solution of $E_{\mbox{\tiny $(D,\bX)$}}\left[\bPsi(\beta_0,\cdots,\beta_{k-1})\right]=0$  with the expectation taken under the true joint distribution of $(D,\bX)$, and $\beta_j^{\ast}=\lim_{n\rightarrow \infty}\betahat_j$.
\end{remark}

\subsection{Accuracy measures and IncV metrics}\label{subsec:metrics}

The AUC and AP can be defined on any risk score  $r(\bX)$ since they are rank-based. The ROC curve is a curve of the true positive rate (TPR) versus the false positive rate (FPR).  Given a cut-off value $c$, the TPR and FPR are the proportions of higher risk $r(\bX) > c$ among the events and non-events respectively, i.e., $\tpr(c) = Pr\left[r(\bX) > c \mid D=1\right]$ and $\fpr(c)=Pr\left[r(\bX) >c \mid D=0\right]$. The AUC can be interpreted as the conditional probability that given a pair of an event and a non-event, the event is assigned with a higher risk score, i.e., $\auc = Pr[r(\bX_i) > r(\bX_j) \mid D_i=1,D_j=0]$. 

The PR curve is a curve of the positive predicted value (PPV) versus the TPR. The PPV is defined as $\ppv(c) = Pr[D=1\mid r(\bX) >c]$, the proportion of subjects with higher risk scores that are events. The AP can be expressed as $\ap = E\left[\ppv\left(r_1(\bX)\right) \right]$ \citep{yuan2018threshold}, where $r_1(\bX)$ denotes the risk score of an event, and the expectation is taken under the distribution of $r_1(\bX)$.  

Both the AUC and AP are proper scoring rules: the true model has the maximum AUC and AP among all the models. One of their differences is that the AP is event-rate dependent \citep{yuan2018threshold}, whereas the AUC does not depend on $\pi$ since it is conditional on the event status.

Let $\Psi_{old}$ and $\Psi_{new}$ denote an accuracy measure $\Psi$ (e.g., AUC or AP) of the existing and new risk models, respectively.  The IncV parameter is defined as $\Delta{\Psi}=\Psi_{new}-\Psi_{old}$, which quantifies the change in $\Psi$ when comparing the new model with the existing one.

\subsection{A data example}\label{subsec:data-result}

Accurate ovarian failure (AOF) is a complication from acute toxicity of exposure to radiation and chemotherapy.  It is defined as permanent loss of ovarian function within 5 years of a cancer diagnosis or no menarche after cancer treatment by age 18. About 6\% of female childhood cancer survivors have AOF.  We evaluate and compare two recently published risk models \citep{clark2020predicting} that predict AOF on an external validation dataset, the St. Jude Lifetime Cohort \citep{hudson2011prospective}, which consists of 875 survivors with 50 AOF events. 

Both models include the following risk factors: age at cancer diagnosis,  cumulative dose of alkylating drugs measured using the cyclophosphamide-equivalent dose, haematopoietic stem-cell transplant, and radiation exposure. The difference between the two models is in the measurement of radiation exposure. The \textit{prescribed-dose} model uses the prescribed radiation doses to the abdominal and pelvic regions, which are routinely available in medical charts. The \textit{ovarian-dose} model uses the minimum of the organ-specific radiation exposure for both ovaries estimated by radiation oncologists.  The equation for calculating the AOF risk using each model is given in the supplementary material of Clark et al. (2020). 

\begin{figure}[h!]
\includegraphics[width=1\textwidth]{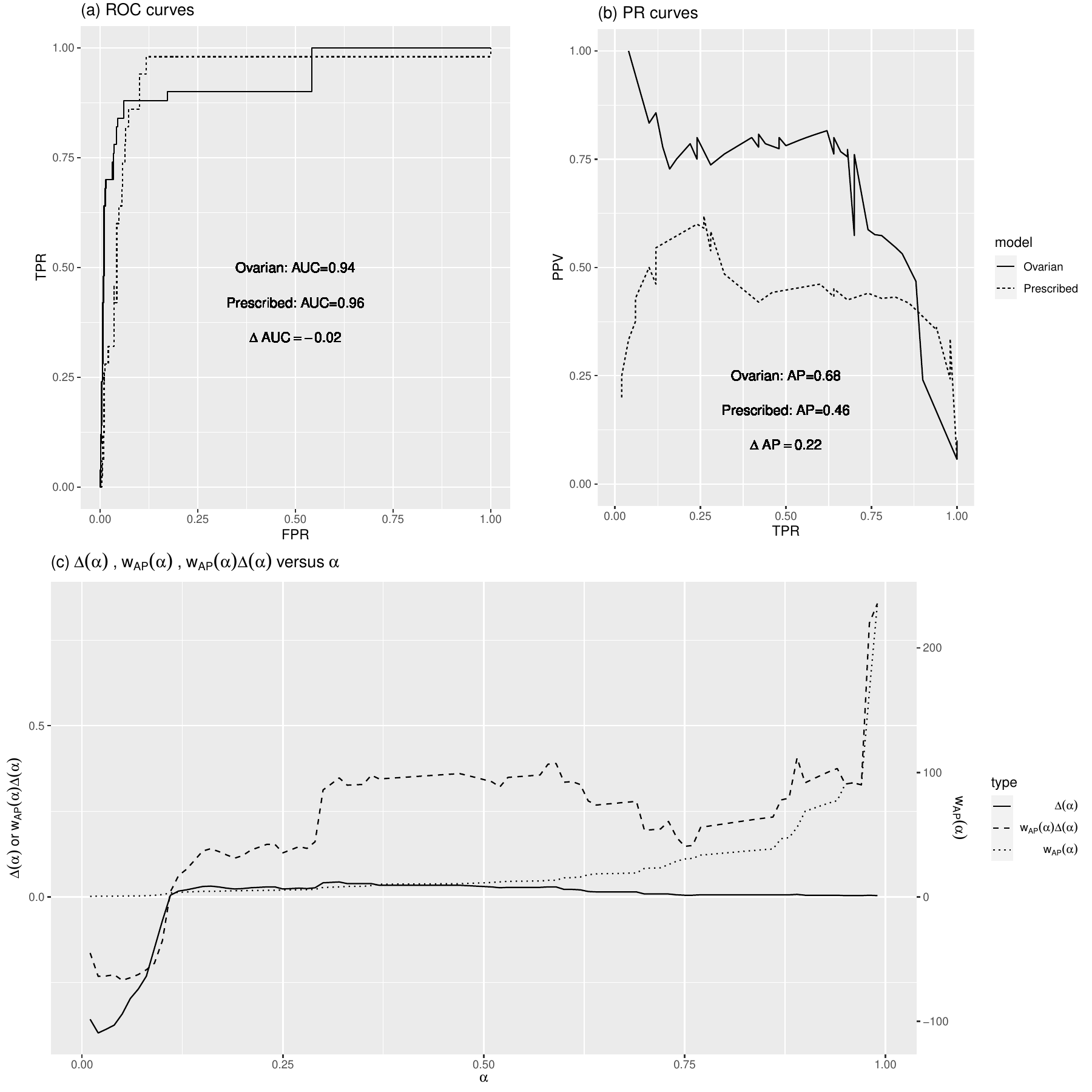}
\caption{Data example: ovarian-dose vs prescribed-dose.} \label{fig:AOF}
\end{figure}

Figure \ref{fig:AOF} (a) shows the ROC curves and PR curves of these two models. The estimated AUC is 0.96 for the prescribed-dose model and 0.94 for the ovarian-dose model; $\Delta \auc$ is estimated to be $-0.02$. The estimated AP is 0.46 for the prescribed-dose model and 0.68 for the ovarian-dose model. The estimated $\Delta \ap$ is $0.22$. The estimation procedure is explained in Appendix.

Based on the IncV in AUC, we conclude that the ovarian-dose model is slightly worse than the prescribed-dose model, and as a result, there is no need to obtain the more expensive ovary dosimetry.  However, based on the IncV in AP, the ovarian-dose model clearly outperforms the prescribed-dose model. Why do these two metrics give conflicting conclusions?

\section{AUC and AP: Measuring the separation of the risk scores between events and non-events} \label{sec:separation}

To answer this question, we first investigate the connections and differences between the AUC and AP using the following three hypothetical risk scores: $r_1$, $r_2$, and $r_3$. We assume that all  the risk scores among non-events follow a standard normal distribution, i.e., $r_j\mid D=0 \sim N(0,1)$, for $j=1,2,3$. However, their distributions among events are different: (i) $r_1 \mid D=1 \sim N(1.8,2)$, (ii) $r_2 \mid D=1\sim N(1.5,1.5)$, and (iii) $r_3\mid D=1 \sim N(3,1.5)$. 

\begin{figure}[h!]
\includegraphics[width=1\textwidth]{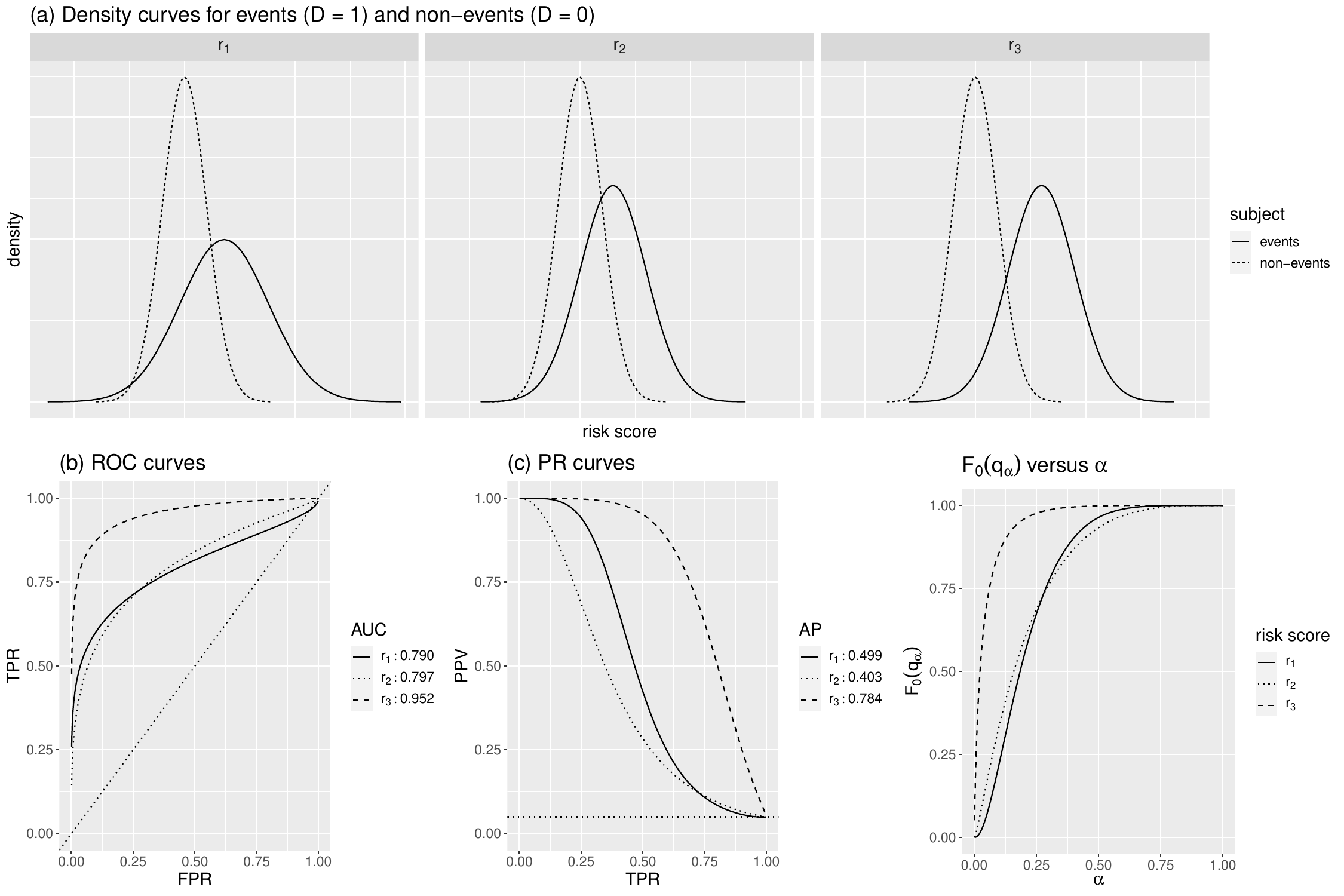}
\caption{Comparison of three hypothetical risk scores $r_1$, $r_2$, and $r_3$ at event rate $\pi=0.05$.}\label{fig:illustration-1}
\end{figure}

Figure \ref{fig:illustration-1} presents the comparisons of these three risk scores under an event rate $\pi=0.05$. Figure \ref{fig:illustration-1} (a) shows their density curves stratified by events and non-events. Among them, the two density curves of $r_3$ are the most separated. Thus, the ROC and PR curves of $r_3$ dominate those of $r_1$ and $r_2$ (Figure \ref{fig:illustration-1} (b)), and consequently, $r_3$ has the largest AUC and AP. In contrast, the ROC and PR curves of $r_1$ and $r_2$ cross: $r_2$ has a slightly larger AUC with $\auc_{r_2} - \auc_{r_1} = 0.007$, but $r_1$ has a considerably larger AP with $\ap_{r_1} - \ap_{r_2} = 0.096$. Figure \ref{fig:illustration-2} exhibits the comparisons between $r_1$ and $r_2$ for three different event rates $\pi=0.2$, $0.05$, and $0.01$.

Analytically, both the AUC and AP measure the separation of the risk score distributions between events and non-events.  Let $F_1(\cdot)$ and $F_0(\cdot)$ denote the cumulative distribution functions (CDFs) of a risk score $r(\bX)$ conditional on $D=1$ (events) and $D=0$ (non-events), respectively. Let $q_{\alpha}=F_1^{-1}(\alpha)$ denote the $\alpha$-th quantile for the distribution $F_1$,  $0\leq \alpha \leq 1$. As shown in equations (\ref{equ:auc-int}) and (\ref{equ:ap-int}) of Appendix, 
the AUC and AP can be expressed as functions of $F_0(q_{\alpha})$, the proportion of non-events whose risk scores are below the $\alpha$-th quantile of the risk scores among events. The $F_0(q_{\alpha})$  measures the separation of the two distributions $F_1$ and $F_0$: the larger the $F_0(q_{\alpha})$ is at a given $\alpha$, the more non-events having lower risk scores, indicating a further separation between these two distributions. For example, the $F_0(q_{\alpha})$ curve of $r_3$ dominates those of $r_1$ and $r_2$ (Figure \ref{fig:illustration-1} (b)), which is consistent with the fact that  $r_3$ has the best separation between events and non-events (Figure \ref{fig:illustration-1} (a)).

Furthermore, we can express $\Delta \auc$ and $\Delta \ap$ as 
\begin{equation}\label{equ:Delta-Psi}
\Delta{\Psi} = \int_0^1 w_{\Psi}(\alpha)\Delta(\alpha) d\alpha,\, \Psi=\auc\, \text{or}\, \ap,
\end{equation}
where  $w_{\Psi}(\alpha)$ is a weight function, and $\Delta(\alpha) = F_{new,0}(q_{new,\alpha}) - F_{old,0}(q_{old,\alpha})$, capturing how much the new working risk model changes the separation of these two distributions at a given $\alpha$. Note that $\Delta(\alpha)$ is independent of $\pi$ because it is conditional on the event outcome. Thus, $\Delta \auc$ and $\Delta \ap$ are weighted averages of $\Delta(\alpha)$, but their weights are different. For $\Delta \auc$, $w_{\auc}(\alpha) \equiv 1$ for $0\leq \alpha \leq 1$, i.e., $\Delta(\alpha)$ is \textit{equally} weighted. For $\Delta \ap$, $w_{\ap}(\alpha)$  is a function of $\alpha$ and $\pi$ (equation (\ref{equ:w_alpha}) of Appendix). 

To visualize how $w_{\ap}(\alpha)$ changes with $\alpha$ and $\pi$,  we plot the $w_{\ap}(\alpha)$ in a log scale against $\alpha$ for different $\pi$ in Figure \ref{fig:illustration-2} (a), in the context of comparing the hypothetical risk scores $r_1$ and $r_2$. For any $\pi$,  $w_{\ap}(\alpha)$ increases with $\alpha$. This tells us that $\Delta \ap$ assigns heavier weights to the upper-tail quantiles of the risk score, representing the higher-risk group, and lighter weights to the lower-tail quantiles, representing the lower-risk group, i.e., $\Delta \ap$ emphasizes the change of the separation in the high-risk group. However, the change is equally weighted in $\Delta \auc$ since $w_{\auc}(\alpha)\equiv 1$. 

Thus,  when the main objective is to identify the high-risk group for intervention, such as when screening for a rare outcome, $\Delta \ap$ is more appropriate than $\Delta \auc$. On the other hand, for some classification problems, such as classifying  early-stage or late-stage  cancer, both the low-risk group and high-risk group may be of interest, and $\Delta \auc$ serves this purpose better. 

Additionally, $w_{\ap}$ is affected by $\pi$. When $\pi$ is smaller, $w_{\ap}(\alpha)$ is larger for $\alpha$ values close to 1 but smaller for $\alpha$ values close to 0. This indicates that, at a lower event rate, if a risk model can better separate the two risk score distributions at the upper quantiles, it will be rewarded more; if it has a worse separation at the lower quantiles, it will be penalized less.

\subsection{Hypothetic risk scores $r_1$ and $r_2$ revisited} \label{sec:r1_r2}

\begin{figure}[h!]
\includegraphics[width=1\textwidth]{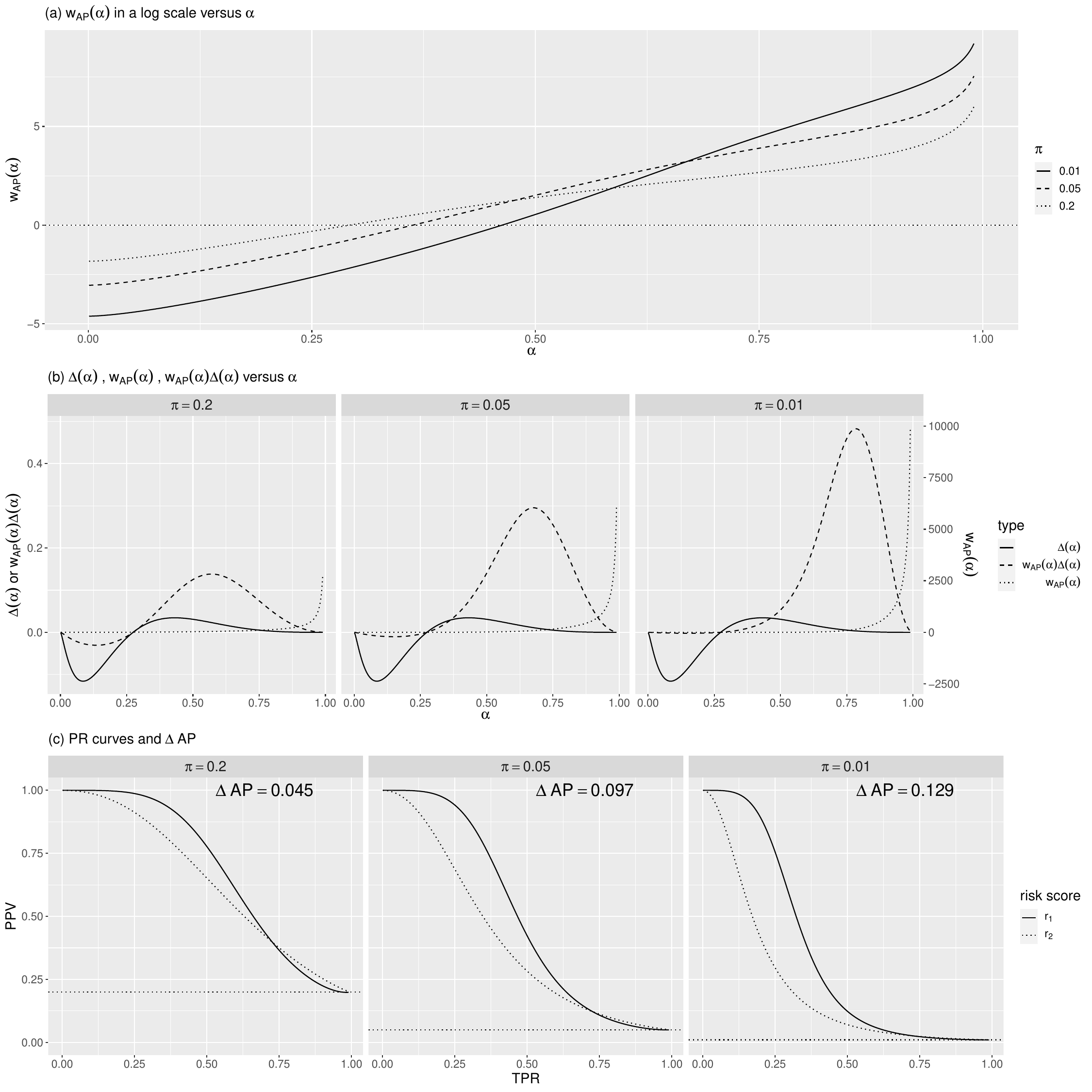}
\caption{Comparison of hypothetical risk scores $r_1$ and $r_2$ under event rates $\pi=0.2, 0.05, 0.01$.}\label{fig:illustration-2}
\end{figure}

Assuming that $r_2$ is from an existing risk model and $r_1$ is from a new one, $\Delta(\alpha)=F_{r_1,0}(q_{r_1,\alpha}) - F_{r_2,0}(q_{r_2,\alpha})$, $\Delta\auc = \auc_{r_1} - \auc_{r_2}$, and $\Delta \ap = \ap_{r_1}-\ap_{r_2}$.  As shown in Figure \ref{fig:illustration-2} (b), $\Delta(\alpha) > 0$ for large $\alpha$, and $\Delta(\alpha)<0$ for small $\alpha$. It indicates that compared to $r_2$, $r_1$ has a better separation for the upper quantiles of the risk score but worse for the lower quantiles.  With equal weighting, $\Delta \auc$ is equivalent to the area under $\Delta(\alpha)$ curve over its entire range. Since the area above 0 is approximately the same as the area below 0, $\Delta \auc \approx 0$.  As mentioned earlier, $\Delta \auc$ is invariant for different $\pi$. Thus, $\Delta \auc = -0.007$ (Figure \ref{fig:illustration-1} (b)) for all three $\pi$ values.

For $\Delta \ap$, the $r_1$'s upper-tail better performance is weighted more than its lower-tail worse performance, which explains $\Delta_{\ap}$ is all positive for the three $\pi$ values (Figure \ref{fig:illustration-2} (c)). Additionally, when $\pi$ gets smaller, the better separation of $r_1$ at the upper quantiles is rewarded more, and meanwhile,  its worse separation at the lower quantiles is penalized less.  Thus, even though $\Delta(\alpha)$ stays the same across different $\pi$, $\Delta{\ap}$ increases as $\pi$ decreases (Figure \ref{fig:illustration-2} (c)). 

\subsection{Data example revisited}\label{subsec:data-revisit}

Let $\Delta(\alpha)=F_{\text{ovarian},0}(q_{ovarian,\alpha}) - F_{\text{prescribed},0}(q_{prescribed,\alpha})$. Figure \ref{fig:AOF} (b) plots the estimated $\Delta(\alpha)$, $w_{\ap}(\alpha)$, and $w_{\ap}(\alpha)\Delta(\alpha)$. It shows that the estimated $\Delta(\alpha)>0$ for $\alpha > 10$\%, whereas the prescribed-dose model performs better with the estimated $\Delta(\alpha)<0$ when $\alpha < 10$\%. It suggests that compared to the prescribed-dose model, the ovarian-dose model separates the events and non-events better among individuals predicted to be at a higher risk.   Overall, under the estimated $\Delta(\alpha)$ curve, the area below zero is slightly larger than the area above zero.  Thus,  the estimated $\Delta \auc$ is negative but close to zero. This indicates that these two models have comparable performance in terms of discrimination. 

However,  the estimated $\Delta \ap$ rewards the superior performance of the ovarian-dose model at the upper quantiles with large weights, and thus, it is positive and sizable. Clark et al. (2020) created four risk groups: low ($<$ 5\%), medium-low (5\% to $<$20\%), medium (20\% to $<$50\%), and high risk ($\geq$ 50\%). The ovarian-dose model classifies 37 individuals (out of 875) as high risk, among which 30 (81\%) experienced AOF, while the prescribed-dose model predicted 13 individuals at high risk, with 6 (46\%) having developed AOF.  This again confirms that the ovarian-dose model is better at identifying high-risk individuals. 

\paragraph*{Comparison with Brier score.} Since both the AUC and AP are rank-based, they are semi-proper scoring rules: a misspecified working risk model and the true model can have the same AUC and AP when they rank the subjects' risks in the same order. We decide to compare these two metrics with the Brier score (BrS), the only strictly proper scoring rule. 

The BrS is the expected squared difference between the binary outcome $D$ and the working risk $p(\bX)$, i.e., $\bs = E_{\mbox{\tiny $(D,\bX)$}}\left\{\left[D-p(\bX)\right]^2\right\}$. The BrS is minimized at the true model, i.e., $p(\bX)=\pi(\bX)$. A non-informative model, assigning the event rate to every subject, i.e., $p(\bX) \equiv \pi$, leads to the maximum BrS value $\pi(1-\pi)$.  A \textit{scaled Brier score} (sBrS) is defined as $\sbs = 1-\bs/[\pi(1-\pi)]$, ranging from 0 and 1, with larger values indicating better performance \citep{steyerberg2010assessing}.

\begin{remark}
Although the BrS cannot be directly expressed as a function of $F_0(q_{\alpha})$, it is closely related to the two distributions $F_1$ and $F_0$.  Specifically, it can be written as 
\begin{align*}
\bs & = E\left\{\left[1-p(\bX)\right]^2\mid D=1\right\}\pi + E\left\{\left[p(\bX)\right]^2\mid D=0\right\}(1-\pi).
\end{align*}
The first expectation is the mean squared prediction error (MSPE) of the working risk $p(\bX)$ for events, determined by the distribution $F_1$, whereas the second expectation is the MSPE for non-events, determined by the distribution $F_0$. Both MSPEs can be expressed as the sum of the variance of $p(\bX)$ and its squared bias from 1 for events and 0 for non-events. A smaller BrS can result from one, or a combination, of the following: (i) the mean of $p(\bX)$ for events closer to 1, (ii) the mean of $p(\bX)$ for non-events  closer to 0, (iii) less variation in $p(\bX)$ for events or non-events or both. All of these lead to a further separation of the two distributions: $F_1$ and $F_0$.
\end{remark}

Let $\Delta \sbs$ denoted the IncV in sBrS. The sBrS is estimated to be 0.23 for the prescribed-dose model and 0.50 for the ovarian-dose model, and $\Delta \sbs$ is estimated to be 0.27. Thus, similar to $\Delta \ap$,  $\Delta \sbs$ favors the ovarian-dose model. 

Why are $\Delta \sbs$ and $\Delta \ap$ consistent in this example? Figure S1 of the supplementary material shows the histogram of the predicted risk $\widehat p_i$ from each model among the AOF and non-AOF individuals.  For the non-AOF individuals, the risk score distributions of these two models are similar. Consequently, the mean and variance of $\widehat p_{i}$ for both models are also similar: the mean is 0.033 for the ovarian-dose model and 0.042 for the prescribed-dose model; their variances are both about 0.0053. The MSPE for the ovarian-dose model is 0.0064, slightly lower than 0.0071 for the prescribed-dose model.

For the AOF events, the risk score distribution of the ovarian model has a heavier right tail. This indicates that the ovarian-dose model pushes more AOF events to the high-risk group. As a result, the mean of $\widehat p_{i}$ for the ovarian-dose model is 0.48, much closer to 1 than 0.23 for the prescribed-dose model. The variance is 0.10 for the ovarian dose model and 0.023 for the prescribed-dose model. The MSPE of the ovarian-dose model is 0.367, much smaller than 0.613 of the prescribed-dose model. Combining the MSPEs for events and non-events weighted by their respective proportions, the estimated BrS for the ovarian-dose model is 0.027, which is smaller than 0.042, the estimated BrS for the prescribed-dose model.

This data example illustrates a comparison of the three IncV parameters: $\Delta \auc$, $\Delta \ap$, and $\Delta \sbs$. Next, we expand the comparison via a numerical study.

\section{Numerical Study}  \label{sec:numerical}

As we are interested in the IncV parameters of the population working risk, not in the IncV estimates from a sample, we do not use simulation studies. In this section, we conduct a numerical study to evaluate the IncV of adding a marker, denoted by $Y$, to a model with an existing marker, denoted by $X$. The IncV parameters can be directly derived from the distributional assumptions described below.  

\subsection{Setting} \label{subsec:assumptions}
Let the markers $X$ and $Y$ be independent standard normal random variables. Given the values of these two markers, a binary outcome $D$ follows a Bernoulli distribution with the probability of $D=1$ via the following model:
\begin{equation}\label{equ:true-model}
\pi(X,Y) =  Pr(D=1\mid X,Y) = \Phi(\beta_0+\beta_1X+\beta_2Y+\beta_3XY),
\end{equation}
where $\Phi(\cdot)$ is the CDF of a standard normal distribution. Given $X$ and $Y$, $\pi(X,Y)$ is the \textit{true} risk. The true model in equation (\ref{equ:true-model}) includes an interaction between $X$ and $Y$, indicating the effect of $X$ on the risk changes with the value of $Y$, and vice versa. 

Typically, in practice, none of the working models are the true model. Having this in mind, we compare the following two misspecified working models: (i) \textit{one-marker model}: $p(X) = \Phi(\gamma_0+\gamma_1X)$, and (ii) \textit{two-marker model}: $p(X,Y) = \Phi(\gamma_0+\gamma_1X+\gamma_2Y)$. 

Here, we consider different values of $\beta_1$, $\beta_2$, $\beta_3$ and $\pi$: $\beta_1=0.3,0.4,\cdots,0.9,1$, $\beta_2 = 0.3,0.4,\cdots,0.9,1$, $\beta_3 = -0.5,-0.4, \cdots,-0.1,0.1,\cdots,0.4,0.5$ (excluding 0), and $\pi=0.01$, $0.05$, $0.1$, $0.2$, $0.5$. Each combination of $(\beta_1,\beta_2,\beta_3,\pi)$ values is referred to as a scenario. Given a scenario, the value of $\beta_0$ can be derived. In the supplementary material, we explain how to obtain the value of $\beta_0$ and calculate the AUC, AP, and sBrS of the one-marker and two-marker models as well as the IncV parameter.

\subsection{Results} \label{subsec:results}

We compare the three IncV parameters based on key considerations for a desirable IncV: (1) size and range, and (2) agreement.  A desirable IncV metric should be sensitive to the change in the predictive performance. If a new model improves the prediction accuracy,  an IncV should have a sizable positive value. It should also be able to reflect a performance deterioration with a sizable negative value. If an IncV is often close to 0, we might question its utility in decision-making. We are also interested in the agreement among the three IncV parameters, given that they focus on different aspects of the prediction performance. 

\subsubsection{Size and Range}

\begin{figure}[h!]
\includegraphics[width=1\textwidth]{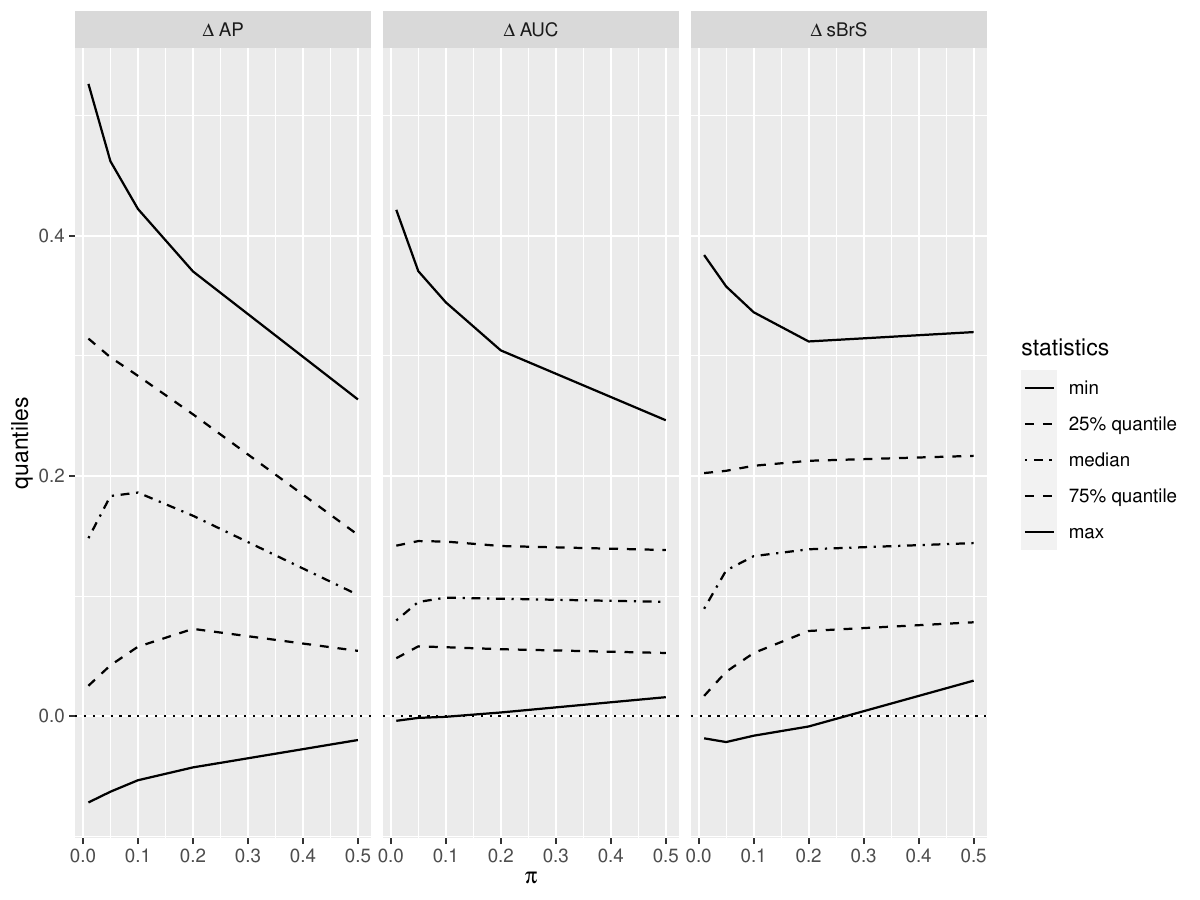}
\caption{Summary statistics of $\Delta \auc$, $\Delta \ap$, and $\Delta \sbs$ versus different event rates $\pi$.}\label{fig:NumIll_quatiles}
\end{figure}

Figure \ref{fig:NumIll_quatiles} plots the summary statistics (minimum, 25\% quantile, median, 75\% quantile, and maximum) of the three IncV metrics under different event rates.  $\Delta \ap$ has the widest range, followed by $\Delta \sbs$, and $\Delta \auc$ has the narrowest range. This difference between $\Delta \auc$ and $\Delta \ap$ is more evident for a lower event rate.  For example, under $\pi=0.01$, the inter-quartile range (IQR) and median of $\Delta \auc$ are both 0.07. In contrast, the IQR of $\Delta \ap$ is much wider, with a range of about 0.41 and a median of $0.21$. 

In addition,  $\Delta \auc$ is negative in less than 1\% of the scenarios (29 out of 3200).  Furthermore, when it is negative, the value is very close to 0, which indicates that $\Delta \auc$ cannot distinguish between a useless marker and a harmful marker \citep{kattan2018index}.  On the other hand, $\Delta \ap$ is negative in about 12\% of the scenarios (389 out of 3200), with a much larger size.

As $\pi$ changes, the range of $\Delta \ap$ varies the most among the three IncV metrics, whereas the quartiles of $\Delta \auc$ remain almost constant. As $\pi$ increases, the ranges of all the IncV metrics get narrower and closer to each other. When $\pi=0.5$, both $\Delta \auc$ and $\Delta \ap$ range from $0.015$ to $0.25$ with a median of $0.089$, and $\Delta \sbs$ ranges from $0.019$ to $0.32$ with a median of $0.12$.

\subsubsection{Agreement}

\paragraph{Correlation.} 
We calculate the Pearson correlation between each pair of the IncV metrics  under each $\pi$ (Table \ref{tab:NumIll_cor}). $\Delta \ap$ and $\Delta \sbs$ are highly correlated for all values of $\pi$. As $\pi$ increases, their correlation decreases from about 1 ($\pi=0.01$) to 0.84 ($\pi=0.5$), but the correlations of $\Delta \auc$ with the other two IncV metrics increases with $\pi$.  When $\pi=0.01$, $\Delta \auc$ and $\Delta \sbs$ are negatively correlated and their correlation $-0.11$ is the smallest among the three pairs; when $\pi=0.5$, they are the highest positively correlated. We also show the scatter plots of each pair under different $\pi$ in Figure S7 (supplementary material).

\paragraph{Concordance.} 
The sign of an IncV metric is often used to decide whether the new model is better than the existing one. Positive IncVs favor the new model, while negative or zero values favor the existing one. Here, we define a concordance measure, which quantifies the consistency of the conclusions reached by a pair of IncV metrics. 

Take $\Delta \ap$ and $\Delta \sbs$ as an example. Under a scenario, we call the pair concordant if both are $>0$ or $\leq 0$. If one is $>0$ and the other is $\leq 0$, the pair is discordant. The measure of concordance is defined as the proportion of scenarios where the pair is concordant minus the proportion of scenarios where it is discordant. For instance, when $\pi=0.01$, $\Delta \ap$ and $\Delta \sbs$ are concordant in about 97\% of the total 640 scenarios (i.e., all the combinations of $\beta_1$, $\beta_2$, and $\beta_3$ values at each $\pi$) and discordant in about 3\%. Thus, the concordance measure is 0.93. 

Table \ref{tab:NumIll_cor} reports the concordance for all three pairs of the IncV metrics under each $\pi$.  The results are similar to those above for the Pearson correlation. When $\pi$ is small, such as 0.01, 0.05, and 0.1, $\Delta \ap$ and $\Delta\sbs$ are the most concordant; when $\pi=0.2$ or 0.5,  $\Delta \auc$ and $\Delta \sbs$ are the most concordant. $\Delta \auc$ and $\Delta\ap$ are the least concordant for all values of $\pi$.

When $\pi$ is close to 0.5, the three IncV metrics tend to agree. Using any of them, we would very likely reach the same conclusion about whether the new model is better. However, when the event rate is low, i.e., for a rare outcome, $\Delta \auc$ can be inconsistent with both $\Delta \sbs$ and $\Delta \ap$.

\subsubsection{ $\Delta \auc$ versus $\Delta \ap$ in selected scenarios}\label{subsec:low-incidence}
\begin{figure}[h!]
\includegraphics[width=1\textwidth]{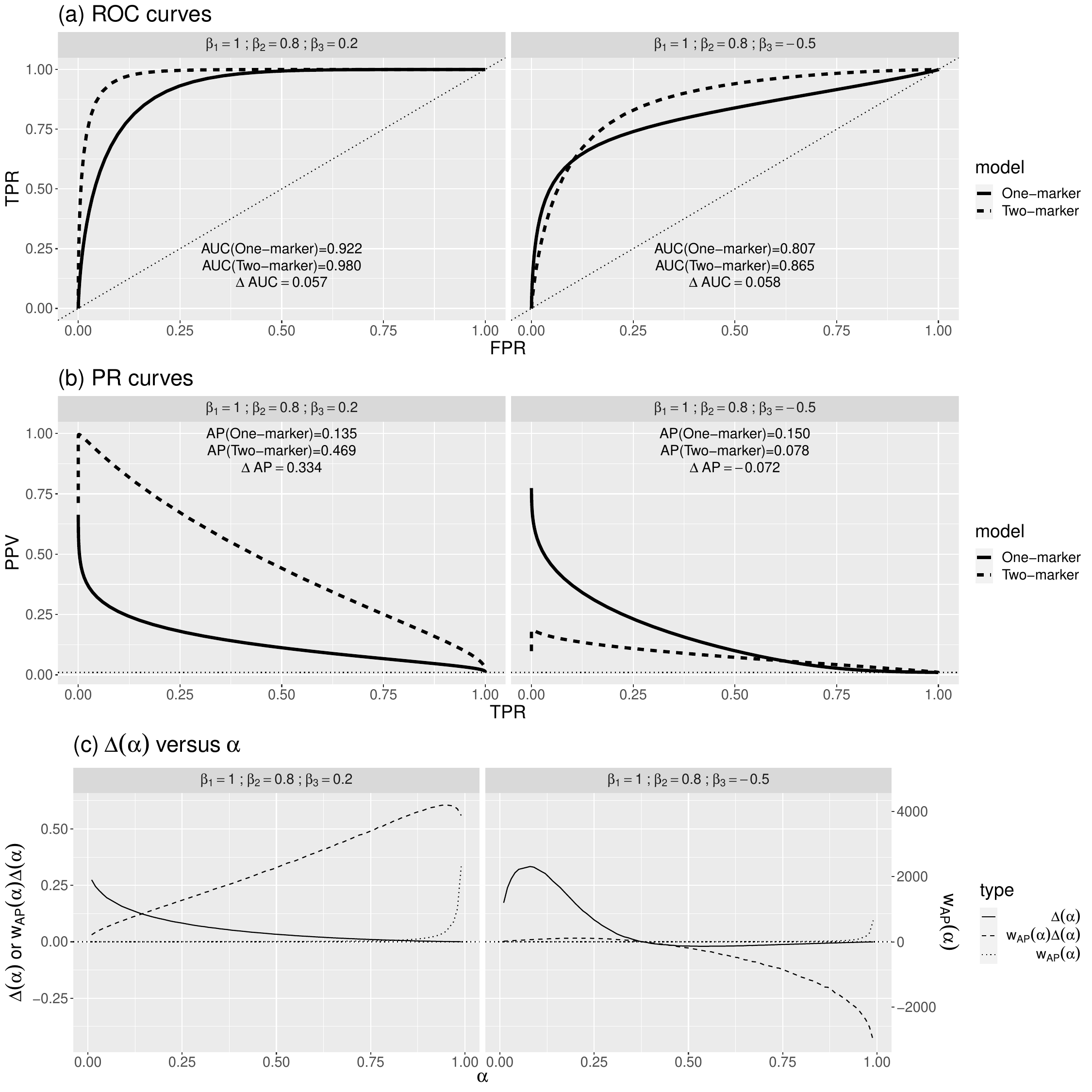}
 \caption{Comparison of two scenarios at event rate $\pi=0.01$: similar $\Delta \auc$ but different $\Delta \ap$.}\label{fig:similar_AUC}
\end{figure}

\begin{figure}[h!]
\includegraphics[width=1\textwidth]{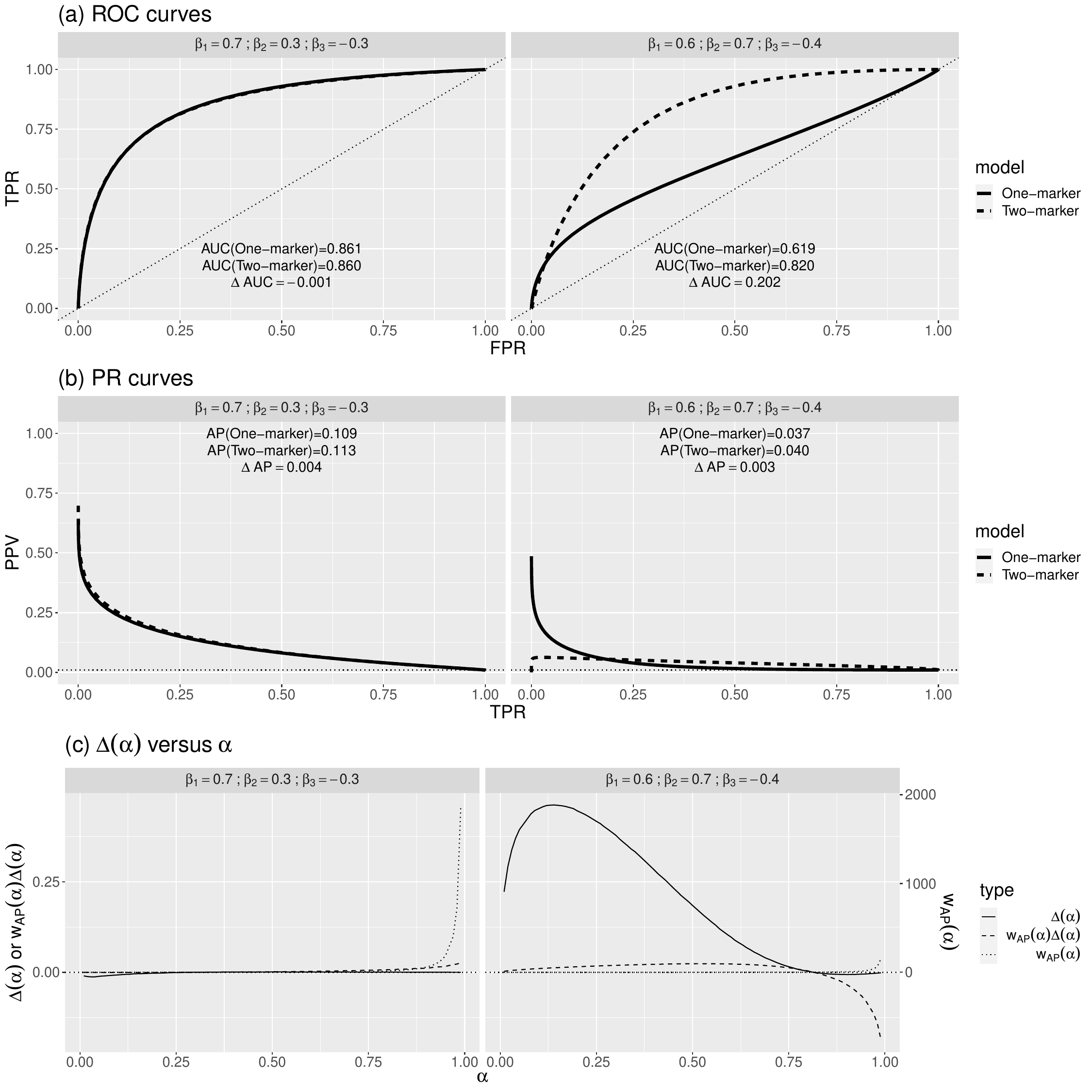}
  \caption{Comparison of two scenarios at event rate $\pi=0.01$: similar $\Delta \ap$ but different $\Delta \auc$.}\label{fig:similar_AP}
\end{figure}

 Next, we single out four scenarios for an in-depth comparison between $\Delta \auc$ and $\Delta \ap$ at $\pi=0.01$. The first two scenarios have similar $\Delta \auc$ but different $\Delta \ap$ (Figure \ref{fig:similar_AUC}),  whereas the next two have similar $\Delta \ap$ but different $\Delta \auc$ (Figure \ref{fig:similar_AP}). 

\paragraph*{Similar $\Delta\auc$ but different $\Delta\ap$.} The two scenarios are (i) $\beta_1=1$, $\beta_2=0.8$, and $\beta_3=0.2$, and (ii) $\beta_1=1$, $\beta_2=0.8$, and $\beta_3=-0.5$. In both cases, $\Delta \auc$ is around 0.06, but $\Delta \ap$ is 0.33 for scenario (i) and -0.072 for scenario (ii). 

In scenario (i), both the ROC and PR curves of the two-marker model dominate those of the one-marker model, respectively. This indicates that the two-marker model is better at each point, and consequently, $\Delta(\alpha)$ is positive throughout (Figure \ref{fig:similar_AUC} (c)). In this case, both $\Delta \auc$ and $\Delta \ap$ are positive. However, the size of $\Delta \ap$ 0.33 is much larger than $\Delta \auc$ of 0.06, due to the large weight $w_{\ap}(\alpha)$ at the upper quantiles (Figure \ref{fig:similar_AUC} (c)).

In scenario (ii), both the two ROC curves and PR curves cross, and $\Delta(\alpha)$ is below zero for upper quantiles and above zero for lower quantiles (Figure \ref{fig:similar_AUC} (c)).   This implies that the two-marker model can better separate between events and non-events for the lower-risk group, but not for the higher-risk group. As a result, $\Delta \auc$ and $\Delta \ap$ are conflicting. $\Delta \auc$ is positive because the area under $\Delta(\alpha)$ curve above zero is larger than that below zero. However, $\Delta \ap$ is negative, as it weights the below-zero $\Delta(\alpha)$ heavily.  In such a situation, is the two-marker model better? The answer should be determined by the objective of the study. If this is a classification problem, the new marker indeed improves the discrimination on average. However, if the goal is to screen for the high-risk group, the two-marker model is worse.

\paragraph*{Similar $\Delta \ap$ but Different $\Delta \auc$.} The next two scenarios are (iii) $\beta_1=0.7$, $\beta_2=0.3$, and $\beta_3=-0.3$, and (iv) $\beta_1=0.6$, $\beta_2=0.7$, and $\beta_3=-0.4$. In both cases, $\Delta \ap$ values are almost 0, but $\Delta \auc$ is approximately 0 for scenario (iii) and 0.202 for scenario (iv).

In scenario (iii), the two ROC curves and the two PR curves are almost identical. This indicates that adding the new marker does not change the separation of the distributions of the risk score between events and non-events. It is also reflected  in Figure \ref{fig:similar_AP} (c) where the entire $\Delta(\alpha)$ curve almost overlaps with the zero line.  Thus, both $\Delta \auc$ and $\Delta \ap$ are close to zero. This is an example of both metrics agreeing that the new marker is ``useless".

In scenario (iv), although the two-marker model makes poorer predictions for the higher-risk group, its prediction is significantly better for the rest. Thus, $\Delta \auc$  is positive and sizable. However, since $\Delta \ap$ weights heavily on the high-risk group, the improvement on the majority is offset by the worse performance at the upper quantiles, which leads to a close-to-zero $\Delta \ap$. If the objective is to identify the high-risk group, the new marker is not helpful because it results in a much lower PPV for subjects with high-risk scores.

\subsection{What if the two-marker model is the true model, i.e., $\beta_3=0$?}\label{subsec:beta3-zero}

Figures S8 and S9 in the supplementary material examine this question and show the scatter plots and plots of the summary statistics of $\Delta \auc$, $\Delta \ap$, and $\Delta \sbs$ for different $\pi$. As expected, all the IncVs are positive. For a smaller $\pi$, $\Delta \ap$ ranges wider than $\Delta \auc$ does. As $\pi$ increases, these two metrics get closer to each other. When $\pi=0.5$, $\Delta \sbs$ has the widest range. 

Since all the IncVs are positive,  their concordance is all 1. Table S1 (supplementary material) lists the Pearson correlation between each pair of the IncV metrics, which are all positive.  When $\pi$ is small, $\Delta \sbs$ is more strongly correlated with $\Delta \ap$ than with $\Delta \auc$. As $\pi$ increases, all three IncV metrics are strongly correlated with each other.

\section{Discussion}\label{sec:discussion}

In this article, we focus on two IncV metrics: $\Delta \auc$ and $\Delta \ap$, for comparing a new risk model with an existing one. We showed that they are intrinsically connected: both can be expressed as a function of $\Delta(\alpha)$, a quantity characterizing the change of the separation of the risk score distributions between events and non-events from an existing risk model to a new one. However,  $\Delta \ap$ emphasizes the change in the higher-risk group, whereas $\Delta \auc$ weights the change equally. Because of this difference, they do not always agree with each other. In the numerical study where both working models are misspecified, the correlation between $\Delta \auc$ and $\Delta \ap$ ranges from negative to positive as the event rate $\pi$ increases from 0.01 to 0.5.  

Additionally, $\Delta \ap$ has a wide range, from negative to positive. However, the size of $\Delta \auc$ values is much smaller, subject to the criticisms of it being insensitive \citep{pepe2004limitations}. $\Delta \auc$ is also rarely negative, unable to reflect the situations when the new model is worse than the existing one. These differences in the magnitude between $\Delta \ap$ and $\Delta \auc$ are more evident when the event rate is low.  The derived analytical expressions of these two metrics provide a new perspective that explains the insensitivity of $\Delta \auc$. When a new model improves the prediction performance for a subgroup, for example, the high-risk group, the AUC weakens the ``local" superior performance by averaging over the entire range. In contrast, the AP augments this ``local" improvement by assigning heavier weights. 

Likewise, if a study objective concerns the low-risk group, we could consider using the area under a curve of negative predicted values versus specificity ($1-$ FPR) as the accuracy metric. Following our derivation of AP, the IncV in this area should also be a weighted average of the change in the separation of the risk score distributions between events and non-events, and the weight is larger for the lower-tail quantiles of the risk score.

In this article, we also find that $\Delta \auc$ and $\Delta \ap$ have different relationships with the only strictly proper scoring rule $\Delta \sbs$. In general, $\Delta \ap$ is always strongly positively correlated with $\Delta \sbs$. On the other hand, as $\pi$ increases, the correlation between $\Delta \auc$ and $\Delta \sbs$  changes from negative to positive, similar to the correlation between $\Delta \auc$ and $\Delta \ap$. 

Our paper focuses on situations when neither of the working models is the true model.  Under such situations, different IncV metrics likely lead to conflicting conclusions on whether the new model is better than the existing one. Choosing which metric to evaluate the new model depends on the purpose of the risk model. Is it for making a diagnosis, screening for subjects with some disease, or identifying the high-risk subjects that will develop an event? The purpose of the risk model requires that it performs well in one or more areas, such as calibration, discrimination, and identification of the high-risk group.  In our data example, the risk models are used to identify the high-risk group that needs counseling for  fertility preservation interventions and hormone replacement therapy. This example illustrates a situation that $\Delta \ap$ is more appropriate whereas $\Delta \auc$ could be misleading. However, in other applications such as classification problems, $\Delta \auc$ would serve better. 

When one of the working models is the true model, Pepe et al. (2013) prove that $H_0: p(X,Y)=p(X)$ is equivalent to the null hypotheses concerning no improvement in the accuracy measures such as the AUC, net reclassification index (NRI), or integrated discrimination improvement (IDI) \citep{pencina2008evaluating}. This is consistent with the results of our numerical study in which the two-marker model is the true model. In this case, the two ROC or PR curves never cross. However, when both working models are misspecified, the two curves might cross, and thus, the above equivalence among the null hypotheses does not hold. 

A good IncV metric should be a proper scoring rule (strictly or semi).  Hilden (2014) pointed out that ``Under no circumstances should a risk assessor, by some clever systematic distortion of the risk assessments, be able to improve his apparent performance."  That is, the true model should always be the winner or among the winners. Although several IncV parameters such as NRI and  IDI has gained popularity, they are not proper scoring rules. Another proper scoring rule is the decision curve and net benefit (NB) \citep{pepe2015net}. We did not include the NB  in our analysis. The NB is designed for making treatment decisions given a patient's risk tolerance, while all three metrics in this paper are about model evaluation.  Also, the NB depends on a threshold probability $p_t$, but the metrics here are threshold-free. As future work, we are interested in comparing $\Delta \auc$, $\Delta \ap$, and $\Delta \sbs$ with $\Delta \text{NB}$.

Because the ranges of AUC, AP, and sBrS are different, the domains of their IncV parameters are also different: $\Delta \auc \in [-0.5,0.5]$, $\Delta \ap \in [\pi-1,1-\pi]$, and $\Delta\sbs \in [-1,1]$. It may be worthwhile to consider rescaling these IncV metrics to range from $-1$ and $1$. Alternatively, an IncV metric can be defined as a ratio such as $\Psi_{new}/\Psi_{old}$.

As mentioned earlier, the AUC is conditional on the disease status. Thus, it can be estimated from either a prospective cohort study or a case-control study. In contrast, the AP is conditional on the risk score, and consequently, it is previously only possible to be estimated from cohort studies.  With the derived expression of the AP in equation (\ref{equ:ap-int}), we provide a potential solution to estimating the AP from a case-control design: one can estimate the AP using (i) an estimated or assumed event rate, and (ii) the risk score distributions of events and non-events estimated from a case-control study.

\numberwithin{equation}{section}

\appendix
\section{Appendix}

\subsection{Estimation of AUC, AP, and sBrS for binary outcomes}\label{append:estimation}

Suppose that the data $\Dfrak=\{(D_i,\bX_i),i=1,\cdots,n\}$ is collected from $n$ subjects. Let $\phat_i$ denoted the estimated risk, described in Remark \ref{rem:true-risk}. Let $\rhat_i$ be a risk score, which is a non-decreasing transformation of $\phat_i$.  The AUC and AP are estimated using $\rhat_i$ by the following nonparametric estimators
$$
\auchat = \frac{\sumin \sumjn I(D_i=1)I(D_j=0)I(\rhat_i>\rhat_j)}{\sumin \sumjn I(D_i=1)I(D_j=0)},
$$
and
$$
\aphat = \frac{\sumin \left[I(D_i=1)\sumjn I(D_j=1)I(\rhat_j>\rhat_i)/\sumjn I(\rhat_j > \rhat_i)\right]}{\sumin I(D_i=1)}.
$$
The BrS can be estimated using $\phat_i$ by $\bshat = n^{-1}\sumin (D_i-\phat_i)^2$. The event rate is estimated as $\pihat = n^{-1}\sum _{i=1}^n D_i$. Then the sBrS is estimated as $\sbshat = 1 - \widehat{\bs}/\left[\pihat(1-\pihat)\right]$.

\subsection{Derivation of AUC and AP}\label{append:AUC-AP-derive}

Let $\pi=Pr(D=1)$ be the event rate, and $r(\bX)=r_{\sbX}$ be a risk score. Let $F(c)=Pr(r_{\sbX}\leq c)$ denote its cumulative distribution function (CDF) for the entire population, and $F_1(c) = Pr(r_{\sbX}\leq c \mid D=1)$ and $F_0(c) = Pr(r_{\sbX}\leq c \mid D=0)$ denote its CDFs for events and non-events, respectively. 

The TPR, FPR, and PPV are 
\begin{align}
\label{equ:TPR} \tpr(c) & = Pr(r_{\sbX} > c \mid D=1) = 1 - F_1(r) \\
\label{equ:FPR} \fpr(c) & = Pr(r_{\sbX}>c \mid D=0) =  1-F_0(r)\\
\nonumber \ppv(c) & = Pr(D=1 \mid r_{\sbX}>c) = \frac{Pr(D=1, r_{\sbX}>c)}{Pr(r_{\sbX}>c)} = \frac{\pi \left[1-F_1(c)\right]}{1-F(c)} \\
 \label{equ:PPV}& = \frac{\pi \left[1-F_1(c)\right]}{\pi \left[1-F_1(c)\right] + (1-\pi)\left[1-F_0(c)\right]}
\end{align}
where $1-F(c) = \pi \left[1-F_1(c)\right] + (1-\pi)\left[1-F_0(c)\right]$. \medskip

\textbf{AUC} is the area under the ROC curve, which can be expressed as 
$$
\auc = \int_{\infty}^{-\infty} \tpr(c) d \fpr(c) = 1 - \int_{\infty}^{-\infty} \fpr(c) d \tpr(c) = \int_{\infty}^{-\infty} \left[1- \fpr(c) \right] d \tpr(c),
$$
because $\int_{\infty}^{-\infty} d \tpr(c) = 1$. Using the expressions in equations (\ref{equ:TPR}) and (\ref{equ:FPR}), we have
$$
\auc = \int_{\infty}^{-\infty} F_0(c) d[1-F_1(c)] = \int_{-\infty}^{\infty} F_0(c)dF_1(c).
$$
Let $q_{\alpha}=F_1^{-1}(\alpha)$ be the $\alpha$-th quantile of the $F_1$ distribution, i.e., $F_1(q_{\alpha})=\alpha$. Thus, let $c=q_{\alpha}$, and we have 
\begin{equation}\label{equ:auc-int}
\auc = \int_0^1 F_0(q_{\alpha}) d\alpha. 
\end{equation}

\textbf{AP} is the area under the PR curve, which can be expressed as
$$
\ap = \int_{\infty}^{-\infty} \ppv(c)d\tpr(c). 
$$
Using the expressions in equations (\ref{equ:TPR}) and (\ref{equ:PPV}), we have
\begin{align*}
\ap & =  \int_{\infty}^{-\infty}  \frac{\pi F_1(c)}{\pi F_1(c) + (1-\pi)F_0(c)} d[1-F_1(c)]  \\
& = \int_{-\infty}^{\infty} \frac{\pi \left[1-F_1(c)\right]}{\pi \left[1-F_1(c)\right] + (1-\pi)\left[1-F_0(c)\right]} dF_1(c)\\
& =  \int_{-\infty}^{\infty} \left\{\frac{\pi \left[1-F_1(c)\right] + (1-\pi)\left[1-F_0(c)\right]}{\pi \left[1-F_1(c)\right]}\right\}^{-1} dF_1(c) \\
& = \int_{-\infty}^{\infty} \left\{1 + \frac{1-\pi}{\pi}\frac{1-F_0(c)}{1-F_1(c)}\right\}^{-1} dF_1(c).
\end{align*}
 Again, let $c=q_{\alpha}$, we have 
\begin{equation}\label{equ:ap-int}
 \ap = \int_{0}^{1} \left\{1+\frac{\pi^{-1}-1}{1-\alpha}\left[1-F_0(q_{\alpha})\right]\right\}^{-1}d\alpha.
 \end{equation}
 
\subsection{Weight $w_{\ap}$ in $\Delta \ap$} \label{append:wAP}
Let $\ap_{old}$ and $\alpha_{new}$ denote the AP of the existing and new models: 
\begin{align*}
\ap_{old} & = \int_{0}^{1} \left\{1+\frac{\pi^{-1}-1}{1-\alpha}\left[1-F_{old,0}(q_{old,\alpha})\right]\right\}^{-1}d\alpha,\\
\ap_{new} & = \int_{0}^{1} \left\{1+\frac{\pi^{-1}-1}{1-\alpha}\left[1-F_{new,0}(q_{new,\alpha})\right]\right\}^{-1}d\alpha.
\end{align*}
Thus, with arithmetic operations, the IncV in AP can be expressed as
\begin{align*}
\Delta \ap  & = \ap_{old} -  \ap_{new}  = \int_{0}^1 w_{\ap}(\alpha) \left[F_{new,0}(q_{new,\alpha}) - F_{old,0}(q_{old,\alpha})\right] d\alpha,
\end{align*}
where
\begin{equation}\label{equ:w_alpha}
w_{\ap}(\alpha) = \frac{\frac{\pi^{-1}-1}{1-\alpha}}{\left[1+(\pi^{-1}-1)\frac{1-F_{new,0}(q_{new,\alpha})}{1-\alpha}\right]\left[1+(\pi^{-1}-1)\frac{1-F_{old,0}(q_{old,\alpha})}{1-\alpha}\right]}.
\end{equation}
It is a function of $\alpha$ and $\pi$. It also depends on $F_{new,0}(q_{new,\alpha})$ and $F_{old,0}(q_{new,\alpha})$. In general, $F_0(q_{\alpha}) \geq \alpha$ because the density curve for non-events is to the left of that for events.  Thus, how the weight changes with $\alpha$ and $\pi$ is mainly determined by the numerator $(\pi^{-1}-1)/(1-\alpha)$. However, when $\pi$ and $\alpha$ are fixed, larger values of $F_{old,0}(r_{old,\alpha})$ or $F_{new,0}(r_{new,\alpha})$ or both, i.e., better performance of at least one model, lead to larger weights.


\section*{Acknowledgements}
The data example is from the St. Jude Lifetime cohort study, supported by National Cancer Institute grant U01CA195547 (PIs Hudson MM and Robinson LL). We thank the St. Jude Lifetime cohort study participants and their families for providing the time and effort for participation and the
internet team at St. Jude Children’s Research Hospital for the development of the web application of the risk prediction models.

\section*{Funding}
Dr. Yuan's research is supported by the Natural Sciences and Engineering Research Council of Canada (RGPIN-2019-04862). 

\section*{Abbreviations}
\textbf{AOF}: acute ovarian failure\\
\textbf{AP}: average positive predicted value or average precision\\
\textbf{AUC}: area under the ROC curve\\
\textbf{BrS}: Brier score\\
\textbf{CDF}: cumulative distribution function\\
\textbf{FPR}: false positive rate\\
\textbf{IDI}: integrated discrimination improvement\\
\textbf{IncV}: incremental value\\
\textbf{NB}: net benefit\\
\textbf{NRI}: net reclassification index\\
\textbf{PPV}: positive predicted value\\
\textbf{PR}: precision-recall\\
\textbf{ROC}: receiver operating characteristic\\
\textbf{sBrS}: scaled Brier score\\
\textbf{TPR}: true positive rate\\

\section*{Availability of data and materials}
The R code for the numerical study and analyzing the data example is available in \href{https://github.com/michellezhou2009/IncVAUCAP}{https://github.com/michellezhou2009/IncVAUCAP}. 

\section*{Ethics approval and consent to participate}
Not applicable.

\section*{Competing interests}
The authors declare that they have no competing interests.

\section*{Authors' contributions}
QZ and YY developed the concept, designed the analytical and numerical studies and drafted the manuscript. LZ conducted the numerical study and analyzed the data example. RB and MH prepared the data from the St. Jude Lifetime cohort Study. LZ, RB, and MH revised the manuscript. All the authors read and approved the final manuscript.


\bibliographystyle{apalike}
\bibliography{AP}

\begin{thebibliography}{}

\bibitem[Badawi et~al., 2018]{badawi2018evaluation}
Badawi, O., Liu, X., Hassan, E., Amelung, P.~J., and Swami, S. (2018).
\newblock Evaluation of icu risk models adapted for use as continuous markers
  of severity of illness throughout the icu stay.
\newblock {\em Critical care medicine}, 46(3):361--367.

\bibitem[Buckley et~al., 2009]{buckley2009c}
Buckley, D.~I., Fu, R., Freeman, M., Rogers, K., and Helfand, M. (2009).
\newblock C-reactive protein as a risk factor for coronary heart disease: a
  systematic review and meta-analyses for the us preventive services task
  force.
\newblock {\em Annals of internal medicine}, 151(7):483--495.

\bibitem[Chaudhury et~al., 2019]{chaudhury2019alzheimer}
Chaudhury, S., Brookes, K.~J., Patel, T., Fallows, A., Guetta-Baranes, T.,
  Turton, J.~C., Guerreiro, R., Bras, J., Hardy, J., Francis, P.~T., et~al.
  (2019).
\newblock Alzheimer's disease polygenic risk score as a predictor of conversion
  from mild-cognitive impairment.
\newblock {\em Translational psychiatry}, 9(1):1--7.

\bibitem[Clark et~al., 2020]{clark2020predicting}
Clark, R.~A., Mostoufi-Moab, S., Yasui, Y., Vu, N.~K., Sklar, C.~A., Motan, T.,
  Brooke, R.~J., Gibson, T.~M., Oeffinger, K.~C., Howell, R.~M., Smith, S.~A.,
  Lu, Z., Robison, L.~L., Chemaitilly, W., Hudson, M.~M., Armstrong, G.~T.,
  Nathan, P.~C., and Yuan, Y. (2020).
\newblock Predicting acute ovarian failure in female survivors of childhood
  cancer: a cohort study in the childhood cancer survivor study (ccss) and the
  st jude lifetime cohort (sjlife).
\newblock {\em The Lancet Oncology}, 21(3):436--445.

\bibitem[Cook et~al., 2006]{cook2006effect}
Cook, N.~R., Buring, J.~E., and Ridker, P.~M. (2006).
\newblock The effect of including c-reactive protein in cardiovascular risk
  prediction models for women.
\newblock {\em Annals of internal medicine}, 145(1):21--29.

\bibitem[Cox et~al., 1972]{cox1972regression}
Cox, D.~R. et~al. (1972).
\newblock Regression models and life tables.
\newblock {\em JR stat soc B}, 34(2):187--220.

\bibitem[Elliott et~al., 2020]{elliott2020predictive}
Elliott, J., Bodinier, B., Bond, T.~A., Chadeau-Hyam, M., Evangelou, E., Moons,
  K.~G., Dehghan, A., Muller, D.~C., Elliott, P., and Tzoulaki, I. (2020).
\newblock Predictive accuracy of a polygenic risk score--enhanced prediction
  model vs a clinical risk score for coronary artery disease.
\newblock {\em Jama}, 323(7):636--645.

\bibitem[Howell et~al., 2019]{howell2019adaptations}
Howell, R.~M., Smith, S.~A., Weathers, R.~E., Kry, S.~F., and Stovall, M.
  (2019).
\newblock Adaptations to a generalized radiation dose reconstruction
  methodology for use in epidemiologic studies: an update from the md anderson
  late effect group.
\newblock {\em Radiation research}, 192(2):169--188.

\bibitem[Hudson et~al., 2011]{hudson2011prospective}
Hudson, M.~M., Ness, K.~K., Nolan, V.~G., Armstrong, G.~T., Green, D.~M.,
  Morris, E.~B., Spunt, S.~L., Metzger, M.~L., Krull, K.~R., Klosky, J.~L.,
  et~al. (2011).
\newblock Prospective medical assessment of adults surviving childhood cancer:
  study design, cohort characteristics, and feasibility of the st. jude
  lifetime cohort study.
\newblock {\em Pediatric blood \& cancer}, 56(5):825--836.

\bibitem[Kattan and Gerds, 2018]{kattan2018index}
Kattan, M.~W. and Gerds, T.~A. (2018).
\newblock The index of prediction accuracy: an intuitive measure useful for
  evaluating risk prediction models.
\newblock {\em Diagnostic and prognostic research}, 2(1):7.

\bibitem[Manning and Sch{\"u}tze, 1999]{manning1999foundations}
Manning, C.~D. and Sch{\"u}tze, H. (1999).
\newblock {\em Foundations of statistical natural language processing}.
\newblock MIT Press, USA.

\bibitem[Mosley et~al., 2020]{mosley2020predictive}
Mosley, J.~D., Gupta, D.~K., Tan, J., Yao, J., Wells, Q.~S., Shaffer, C.~M.,
  Kundu, S., Robinson-Cohen, C., Psaty, B.~M., Rich, S.~S., et~al. (2020).
\newblock Predictive accuracy of a polygenic risk score compared with a
  clinical risk score for incident coronary heart disease.
\newblock {\em Jama}, 323(7):627--635.

\bibitem[Pencina et~al., 2008]{pencina2008evaluating}
Pencina, M.~J., D'Agostino~Sr, R.~B., D'Agostino~Jr, R.~B., and Vasan, R.~S.
  (2008).
\newblock Evaluating the added predictive ability of a new marker: from area
  under the roc curve to reclassification and beyond.
\newblock {\em Statistics in medicine}, 27(2):157--172.

\bibitem[Pepe, 2003]{pepe2003statistical}
Pepe, M.~S. (2003).
\newblock {\em The statistical evaluation of medical tests for classification
  and prediction}.
\newblock Oxford University Press, Oxford.

\bibitem[Pepe et~al., 2015]{pepe2015net}
Pepe, M.~S., Fan, J., Feng, Z., Gerds, T., and Hilden, J. (2015).
\newblock The net reclassification index (nri): a misleading measure of
  prediction improvement even with independent test data sets.
\newblock {\em Statistics in biosciences}, 7(2):282--295.

\bibitem[Pepe et~al., 2004]{pepe2004limitations}
Pepe, M.~S., Janes, H., Longton, G., Leisenring, W., and Newcomb, P. (2004).
\newblock Limitations of the odds ratio in gauging the performance of a
  diagnostic, prognostic, or screening marker.
\newblock {\em American journal of epidemiology}, 159(9):882--890.

\bibitem[Pepe et~al., 2013]{pepe2013testing}
Pepe, M.~S., Kerr, K.~F., Longton, G., and Wang, Z. (2013).
\newblock Testing for improvement in prediction model performance.
\newblock {\em Statistics in medicine}, 32(9):1467--1482.

\bibitem[Raghavan et~al., 1989]{raghavan1989critical}
Raghavan, V., Bollmann, P., and Jung, G.~S. (1989).
\newblock A critical investigation of recall and precision as measures of
  retrieval system performance.
\newblock {\em ACM Transactions on Information Systems (TOIS)}, 7(3):205--229.

\bibitem[Steyerberg et~al., 2010]{steyerberg2010assessing}
Steyerberg, E.~W., Vickers, A.~J., Cook, N.~R., Gerds, T., Gonen, M.,
  Obuchowski, N., Pencina, M.~J., and Kattan, M.~W. (2010).
\newblock Assessing the performance of prediction models: a framework for some
  traditional and novel measures.
\newblock {\em Epidemiology (Cambridge, Mass.)}, 21(1):128.

\bibitem[Su et~al., 2015]{su2015relationship}
Su, W., Yuan, Y., and Zhu, M. (2015).
\newblock A relationship between the average precision and the area under the
  roc curve.
\newblock In {\em Proceedings of the 2015 International Conference on The
  Theory of Information Retrieval}, pages 349--352. ACM.

\bibitem[Tang et~al., 2019]{tang2019prediction}
Tang, M., Hu, P., Wang, C.-F., Yu, C.-Q., Sheng, J., and Ma, S.-J. (2019).
\newblock Prediction model of cardiac risk for dental extraction in elderly
  patients with cardiovascular diseases.
\newblock {\em Gerontology}, 65(6):591--598.

\bibitem[Uno et~al., 2007]{uno2007evaluating}
Uno, H., Cai, T., Tian, L., and Wei, L. (2007).
\newblock Evaluating prediction rules for t-year survivors with censored
  regression models.
\newblock {\em Journal of the American Statistical Association}, 102:527--537.

\bibitem[Xiao et~al., 2019]{xiao2019comparison}
Xiao, J., Ding, R., Xu, X., Guan, H., Feng, X., Sun, T., Zhu, S., and Ye, Z.
  (2019).
\newblock Comparison and development of machine learning tools in the
  prediction of chronic kidney disease progression.
\newblock {\em Journal of translational medicine}, 17(1):119.

\bibitem[Yuan et~al., 2015]{yuan2015threshold}
Yuan, Y., Su, W., and Zhu, M. (2015).
\newblock Threshold-free measures for assessing the performance of medical
  screening tests.
\newblock {\em Frontiers in Public Health}, 3:57.

\bibitem[Yuan et~al., 2018]{yuan2018threshold}
Yuan, Y., Zhou, Q.~M., Li, B., Cai, H., Chow, E.~J., and Armstrong, G.~T.
  (2018).
\newblock A threshold-free summary index of prediction accuracy for censored
  time to event data.
\newblock {\em Statistics in medicine}, 37(10):1671--1681.

\bibitem[Zweig and Campbell, 1993]{zweig1993receiver}
Zweig, M.~H. and Campbell, G. (1993).
\newblock Receiver-operating characteristic (roc) plots: a fundamental
  evaluation tool in clinical medicine.
\newblock {\em Clinical chemistry}, 39(4):561--577.

\end{thebibliography}


\begin{thebibliography}{}

\bibitem[Borchers, 2019]{pracma2019}
Borchers, H.~W. (2019).
\newblock {\em pracma: Practical Numerical Math Functions}.

\bibitem[Hasselman, 2018]{nleqslv2018}
Hasselman, B. (2018).
\newblock {\em nleqslv: Solve Systems of Nonlinear Equations}.

\bibitem[{R Core Team}, 2020]{team2020r}
{R Core Team} (2020).
\newblock {\em R: A language and environment for statistical computing}.
\newblock R Foundation for Statistical Computing.

\end{thebibliography}



\begin{table}[h!]
\centering
\caption{Pearson correlation and concordance measure of each pair of the IncV metrics for different event rates $\pi$.}  
\label{tab:NumIll_cor}
\vspace{0.2cm}
\begin{tabular}{r|r|r|r|r|r}
\hline\hline 
\multicolumn{6}{c}{}\\
\multicolumn{6}{c}{Pearson Correlation}\\
\hline
Comparison & $\pi=0.01$ &  $\pi=0.05$ & $\pi=0.1$ & $\pi=0.2$ & $\pi=0.5$ \\
\hline
$\Delta \sbs$ vs $\Delta \ap$ &  0.995 &  0.992 &  0.986 &  0.971 &  0.837\\
$\Delta \sbs$ vs $\Delta \auc$ & -0.111 &  0.262 &  0.479 &  0.718 &  0.932\\
$\Delta \auc$ vs $\Delta \ap$ & -0.086 &  0.296 &  0.505 &  0.708 &  0.888\\
\hline
\multicolumn{6}{c}{}\\
\multicolumn{6}{c}{Concordance}\\
\hline
Comparison & $\pi=0.01$ &  $\pi=0.05$ & $\pi=0.1$ & $\pi=0.2$ & $\pi=0.5$ \\
\hline
$\Delta \sbs$ vs $\Delta \ap$ & 0.931 & 0.922 & 0.897 & 0.856 & 0.922\\
$\Delta \sbs$ vs $\Delta \auc$ & 0.659 & 0.750 & 0.828 & 0.928 & 1.000\\
$\Delta \auc$ vs $\Delta \ap$ & 0.591 & 0.672 & 0.725 & 0.784 & 0.922\\
\hline\hline
\end{tabular}
\end{table}

\clearpage
\newpage
\section*{Additional Files}
\subsection*{Supplementary material}
The supplementary material includes (i) the histograms of the predicted AOF risk obtained from the prescribed-dose model and ovarian-dose model for individuals with and without AOF, respectively, (ii) the procedure of obtaining the IncV parameters under the distributional assumptions of the numerical study, (iii) the results of the numerical study scenarios in which neither of the working risk models is the true model, including plots of the values of each IncV metric for all the scenarios under different event rates, and the scatter plots of each pair of the IncV metrics, and (iv) the results for the scenarios where the two-marker model is the true model, including plots of the values of each IncV metric for all the scenarios under different event rates, plots of their summary statistics, and a table listing the Pearson correlation of each pair of the IncV metrics. (PDF file)
\end{document}


\title{Supplementary Material of \\
"Is the new model better? One metric says yes, but the other says no. Which metric do I use?"
}
\author{Qian M. Zhou \and Zhe Lu \and Russell J. Brooke \and Melissa M Hudson \and Yan Yuan}
\date{}
\maketitle

\noindent The supplementary material includes 
\begin{itemize}
\item the histograms of the predicted AOF risk for the data example, 
\item the procedure of obtaining the IncV parameters under the distributional assumptions of the numerical study scenarios in which neither of the working risk models is the true model, and
\item the results of the numerical study scenarios in which the two-marker model is the true model.
\end{itemize}

\section{Results of AOF data example}

\begin{figure}[!h]
\centering
\includegraphics[width=1\textwidth]{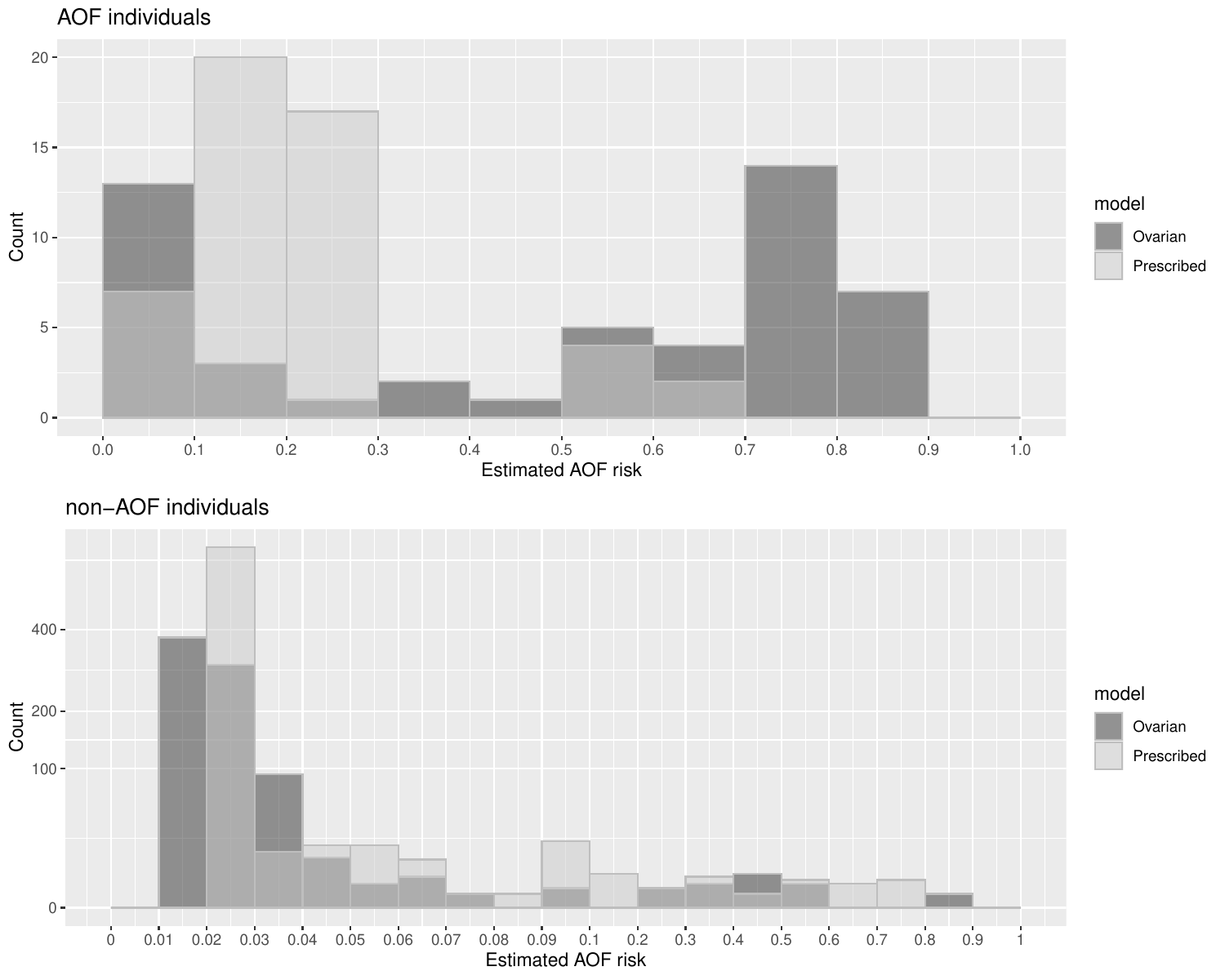}
\caption{Results of Data Example: Histograms of the predicted AOF risk obtained from the prescribed-dose model and ovarian-dose model for individuals with and without AOF, respectively}\label{fig:AOF_density}
\end{figure}

Figure \ref{fig:AOF_density} shows the histograms of the predicted AOF risk obtained from the ovarian-dose model and the prescribed-dose model among individuals with and without AOF, respectively. 

\section{Numerical study: Calculation of IncV parameters}

All the calculations are implemented by the R software \citep{team2020r}. The integrals are calculated by the R package \textit{pracma} \citep{pracma2019}. We use the R function  \textit{uniroot} of the R package \textit{stat} for obtaining the solution to one-dimensional equations, and use the R package  \textit{nleqslv} \citep{nleqslv2018} for solving multi-dimensional equations. 

\medskip

\noindent{\bf Distributional assumptions.} Let $\pi=Pr(D=1)$. The two markers $X$ and $Y$ are independent standard normal distribution. Given $X$ and $Y$, the binary outcome $D$ follows a Bernoulli distribution with the probability of $D=1$ given as
$$
Pr(D=1\mid X,Y) = \Phi(\beta_0 + \beta_1 X + \beta_2 Y + \beta_3 XY) \triangleq \pi(X,Y;\bbeta),
$$
where $\bbeta=(\beta_0,\beta_1,\beta_2,\beta_3)'$, and $\Phi$ is the cumulative distribution function of a standard normal distribution. In the numerical study, we consider two misspecified working risk models: (i) \textit{one-marker model}: $p(X) = \Phi(\gamma_0+\gamma_1X)$, and (ii) \textit{two-marker model}: $p(X,Y) = \Phi(\gamma_0+\gamma_1X+\gamma_2Y)$. 

A combination of the $(\beta_1,\beta_2,\beta_3,\pi)$ values is referred to as a scenario. Given a scenario, we first obtain the value of $\beta_0$. Second, we obtain the regression coefficients  $(\gamma_0,\gamma_1)$ for the one-marker model and  $(\gamma_0,\gamma_1,\gamma_2)$ for the two-marker model. Third, we calculate the accuracy measures of each model. The IncV parameters are calculated by $\Delta \Psi=\Psi_{M_2} - \Psi_{M_1}$, where $\Psi$ denotes an accuracy metric, such as the AUC, AP, and sBrS, $M_1$ denotes the one-marker model, and $M_2$ denotes the two-marker model. \medskip

\noindent {\bf Calculation of $\beta_0$ value.} The event rate $\pi$ can be expressed as
\begin{equation}\label{equ:pi}
\pi = Pr(D=1) = \int_{-\infty}^{\infty}\int_{-\infty}^{\infty} \pi(x,y;\bbeta)\phi(x)\phi(y) dx dy,
\end{equation}
where $\phi(\cdot)$ is the probability density function (PDF) of the standard normal distribution. Given the values of  $(\beta_1,\beta_2,\beta_3,\pi)$, the right hand side of the equation (\ref{equ:pi}) is a function of $\beta_0$. By solving this equation, we can obtain the value of $\beta_0$. \medskip

\noindent{\bf Calculation of regression parameters for working risk models.} We can express the working risk from each model by $\Phi(\bgamma'\bU)$. For the one-marker model, $\bU=(1,X)'$ and $\bgamma=(\gamma_0,\gamma_1)'$; for the two-marker model, $\bU=(1,X,Y)'$ and $\bgamma=(\gamma_0,\gamma_1,\gamma_2)'$. The \textit{population} values (i.e., limiting values) of the regression parameters $\bgamma$ are the solution to the following estimating equation (EE)

\begin{equation}\label{equ:model-EE}
\psi(\bgamma) = E_{X,Y,D}\left\{\bU \frac{\phi(\bgamma'\bU)}{\Phi(\bgamma'\bU)\left[1-\Phi(\bgamma'\bU)\right]} \left[D -\Phi(\bgamma'\bU) \right]\right\} = 0,
\end{equation}
where the expectation is taken with respect to the joint distribution of $(X,Y,D)$. This EE (\ref{equ:model-EE}) can be re-written as
\begin{align*}
\psi(\bgamma) & = E_{X,Y}\left\{\bU \frac{\phi(\bgamma'\bU)}{\Phi(\bgamma'\bU) \left[1-\Phi(\bgamma'\bU)\right]} E\left[D -\Phi(\bgamma'\bU) \mid X,Y \right]\right\}  \\
& = \int_{-\infty}^{\infty}\int_{-\infty}^{\infty}\bu \frac{\phi(\bgamma'\bu)}{\Phi(\bgamma'\bu) \left[1-\Phi(\bgamma'\bu)\right]}\left[\pi(x,y;\bbeta) -\Phi(\bgamma'\bu)  \right] \phi(x)\phi(y) dx dy.
\end{align*}
Given the values of $\bbeta$, the EE is a function of only $\bgamma$. Thus, the solution to this EE gives the population values of $\bgamma$.  \medskip

\noindent {\bf Calculation of the AUC, AP, and sBrS.} Given the values of $\bgamma$, we can obtain the working risk $\Phi(\bgamma'\bU)$ for any given values of $\bU$. The Brier score can be expressed as
$$
\bs = E\left\{\pi(X,Y;\bbeta)\left[1-\pi(X,Y;\bbeta)\right]\right\} + E\left\{\left[\pi(X,Y;\bbeta) - \Phi(\bgamma'\bU)\right]^2\right\},
$$
which can be calculated by
$$
\bs = \int_{-\infty}^{\infty} \int_{-\infty}^{\infty} \left\{\pi(X,Y;\bbeta)\left[1-\pi(X,Y;\bbeta)\right] + \left[\pi(X,Y;\bbeta) - \Phi(\bgamma'\bU)\right]^2 \right\} \phi(x)\phi(y) dx dy.
$$
Thus, $\sbs = 1 - \bs/[\pi(1-\pi)]$.

Both the AUC and AP are invariant under any risk score that is a non-decreasing transformation of the working risk. Thus,  we consider the risk score $r=\bgamma'\bU$.  As shown in Appendix A.1, both the AUC and AP depends on the distribution of $r$ for events and non-events. Let $f(c)$ denote the (unconditional) PDF of $r$. Note that for the one-marker model, $r$ follows a normal distribution with mean $\gamma_0$ and variance $\gamma_1^2$, i.e., $f(c) = \frac{1}{|\gamma_1|}\phi\left(\frac{c-\gamma_0}{\gamma_1}\right)$. For the two-marker model, the PDF of $r$ can be obtained by $f(c)=f_{r\mid X}(c|x)\phi(x)$ or $f_{r\mid Y}(c|y)\phi(y)$, where $f_{r\mid X}(c|x) = \frac{1}{|\gamma_2|}\phi\left(\frac{c-\gamma_0-\gamma_1x}{\gamma_2}\right)$ is the conditional PDF of $r$ given $X=x$,  and $f_{r\mid Y}(c|y) = \frac{1}{|\gamma_1|}\phi\left(\frac{c-\gamma_0-\gamma_2y}{\gamma_1}\right)$ is the conditional PDF of $R$ given $Y=y$.

The PDF of $r$ given $D=1$ is 
\begin{align}
\nonumber f_1(c) & = \frac{\int_{-\infty}^{\infty} Pr(D=1\mid r=c,X=x)f_{r\mid X}(c\mid x) \phi(x) dx}{Pr(D=1)} \\
\label{equ:f1-x}& = \frac{\int_{-\infty}^{\infty}\pi\left(x,\frac{c-\gamma_0-\gamma_1x}{\gamma_2};\bbeta\right)f_{r\mid X}(c\mid x) \phi(x) dx}{\pi}
\end{align}
or 
\begin{align}
\nonumber f_1(c) & = \frac{\int_{-\infty}^{\infty}Pr(D=1\mid r=c,Y=y)f_{r\mid Y}(c\mid y) \phi(y) dy}{Pr(D=1)} \\
\label{equ:f1-y}& = \frac{\int_{-\infty}^{\infty}\pi\left(\frac{c-\gamma_0-\gamma_2 y}{\gamma_1},y;\bbeta\right)f_{r\mid Y}(c\mid y) \phi(y) dy}{\pi}.
\end{align}
Note that for the one-marker model, we can only use the equation (\ref{equ:f1-y}) because $r$ only depends on $X$. Also, $f_{r\mid Y}(c\mid y) = f(c)= \frac{1}{|\gamma_1|}\phi\left(\frac{c-\gamma_0}{\gamma_1}\right)$. Consequently, the CDF of $r$ given $D=1$ is
$F_1(c) = \int_{-\infty}^c f_1(z) dz$.

Similarly, the PDF of $r$ given $D=0$ is
$$
f_0(c) = \frac{\int_{-\infty}^{\infty}\left[1-\pi\left(x,\frac{c-\gamma_0-\gamma_1x}{\gamma_2};\bbeta\right)\right]f_{r\mid X}(c\mid x) \phi(x) dx}{1-\pi}
$$
or
$$
f_0(c) = \frac{\int_{-\infty}^{\infty}\left[1-\pi\left(\frac{c-\gamma_0-\gamma_2 y}{\gamma_1},y;\bbeta\right)\right]f_{r\mid Y}(c\mid y) \phi(y) dy}{1-\pi}.
$$
And the CDF of $r$ given $D=0$ is  $F_0(c) = \int_{-\infty}^c f_0(z) dz$.

The AUC is given as
$$
\auc = \int_{-\infty}^{\infty} F_0(c)f_1(c) dc,
$$
and the AP is given as
$$
\ap = \int_{-\infty}^{\infty} \left[1+\left(\pi^{-1}-1\right)\frac{1-F_0(c)}{1-F_1(c)}\right]^{-1}f_1(c) dc.
$$

\section{Results of the numerical study}

\begin{figure}
\centering
\includegraphics[width=1\textwidth]{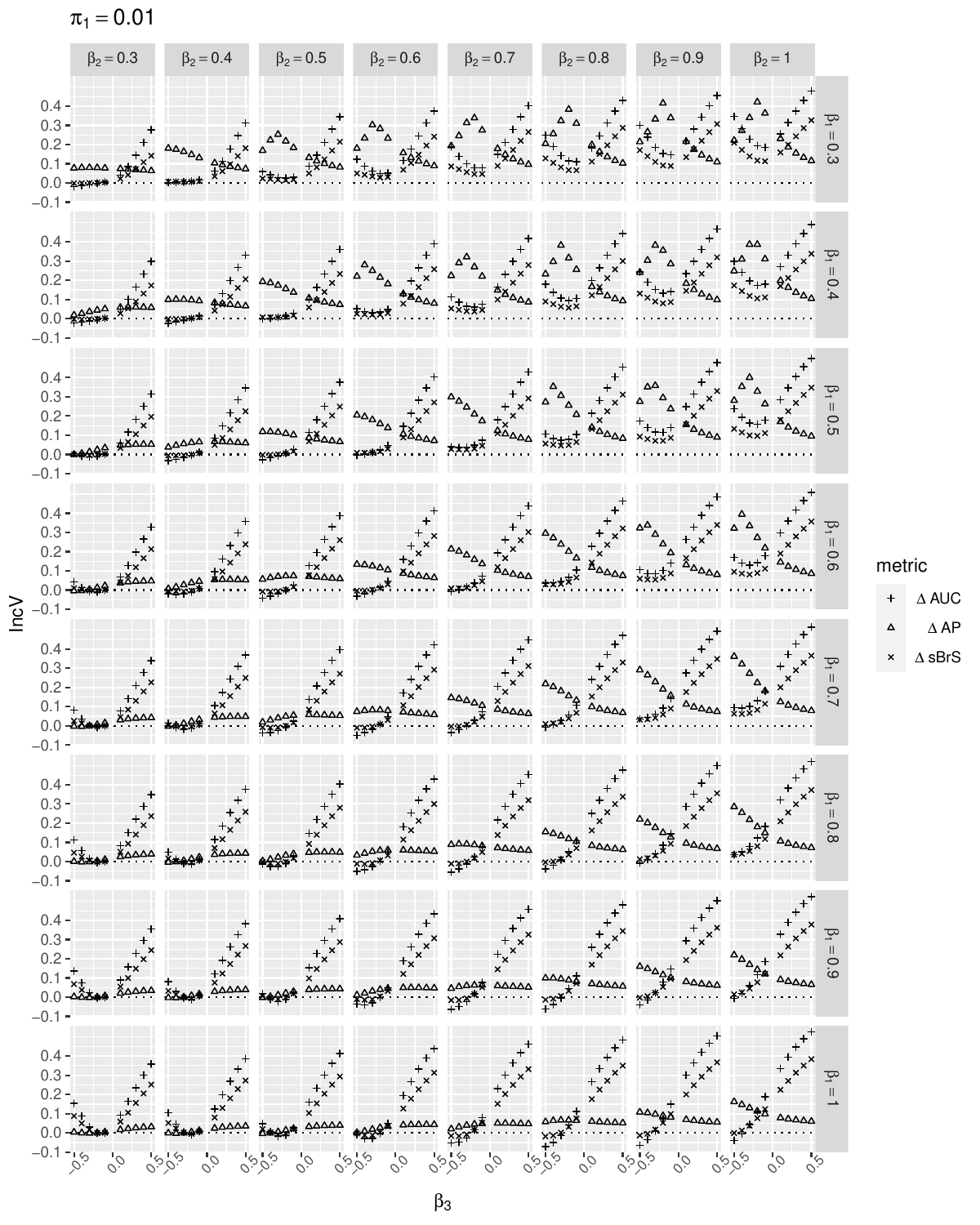}
\caption{The $\Delta \auc$, $\Delta \ap$, and $\Delta \sbs$ under the event rate $\pi=0.01$.}\label{fig:pi-0.01}
\end{figure}

\begin{figure}
\centering
\includegraphics[width=1\textwidth]{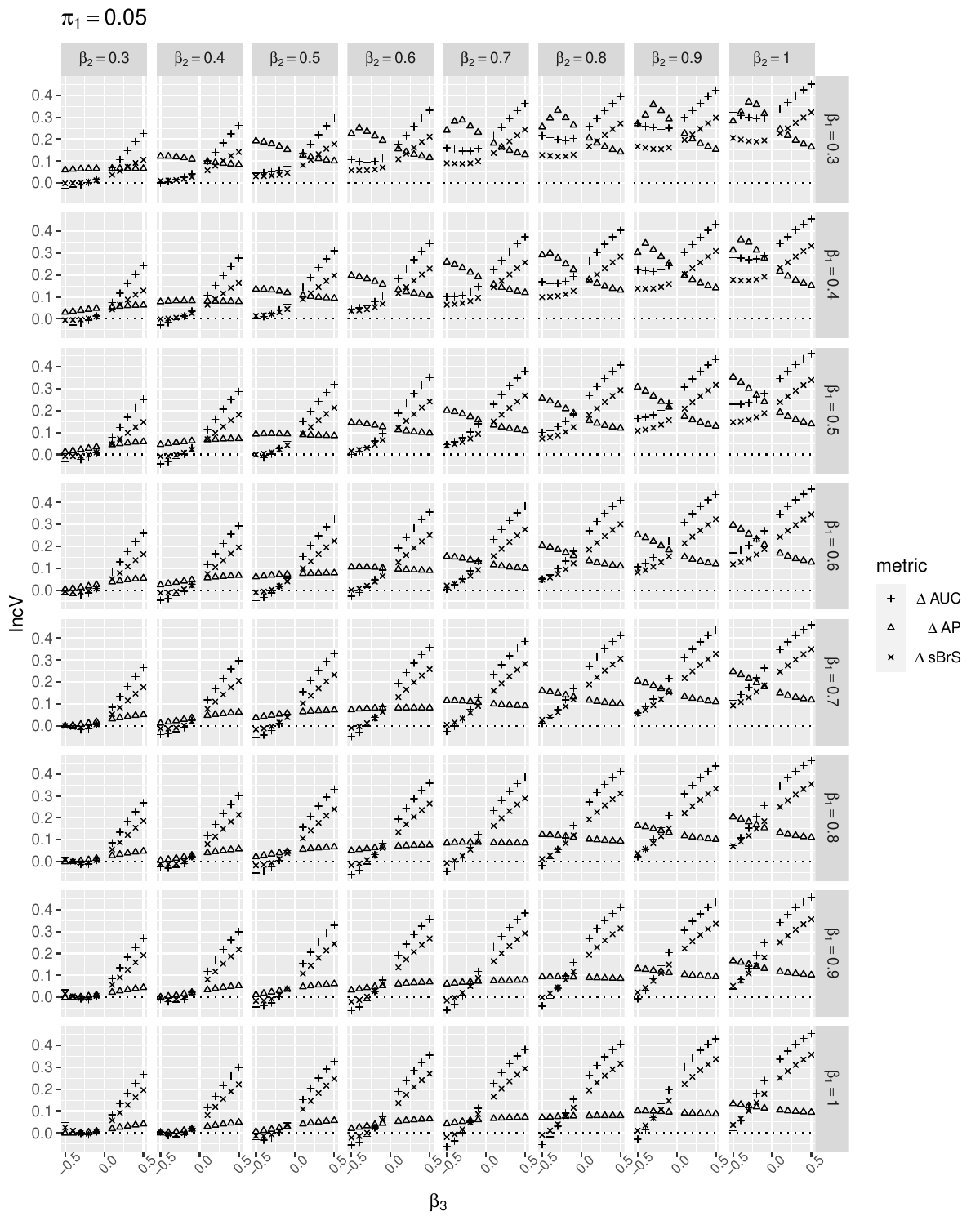}
\caption{The $\Delta \auc$, $\Delta \ap$, and $\Delta \sbs$ under the event rate $\pi=0.05$.}\label{fig:pi-0.05}
\end{figure}

\begin{figure}
\centering
\includegraphics[width=1\textwidth]{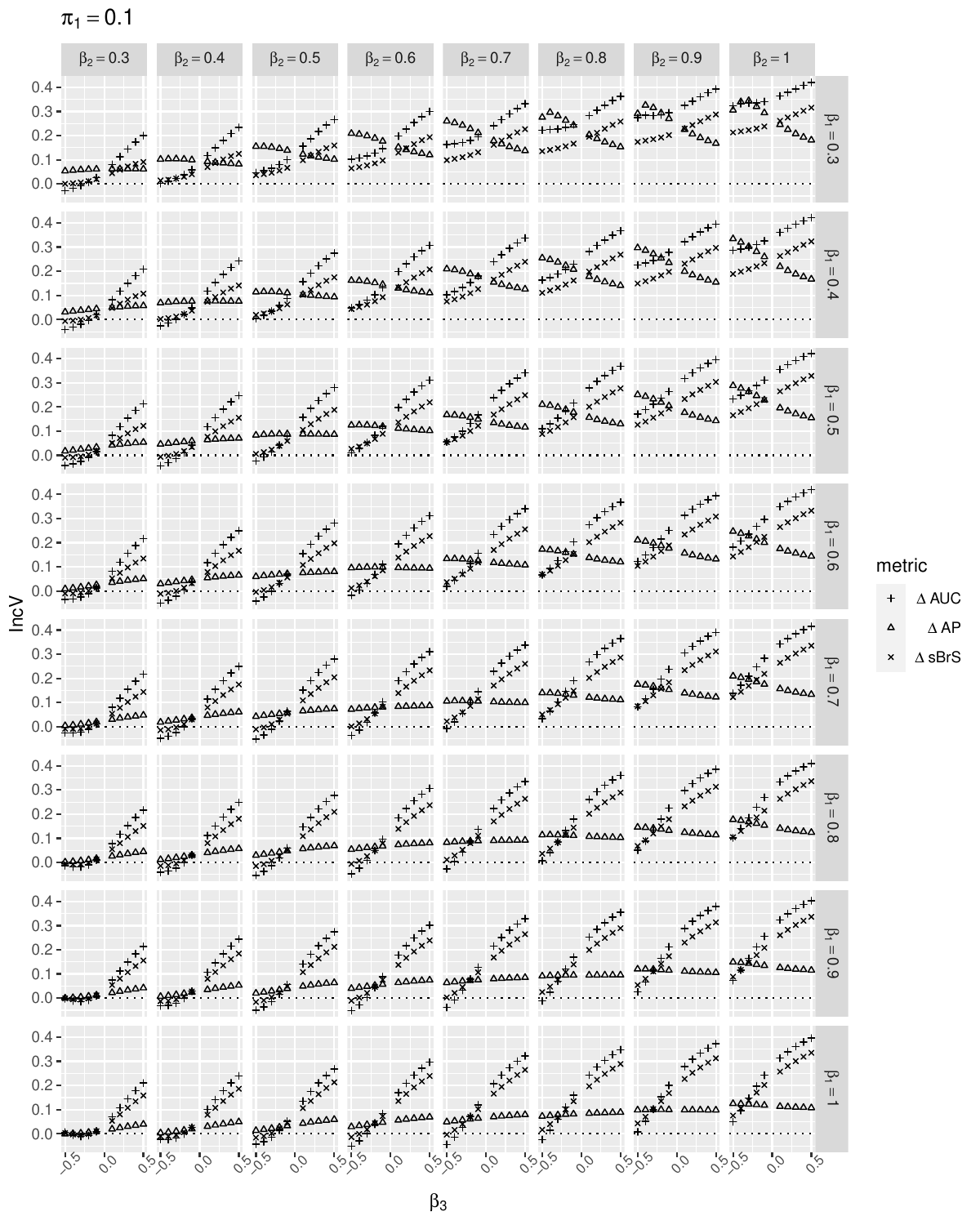}
\caption{The $\Delta \auc$, $\Delta \ap$, and $\Delta \sbs$ under the event rate $\pi=0.1$.}\label{fig:pi-0.1}
\end{figure}

\begin{figure}
\centering
\includegraphics[width=1\textwidth]{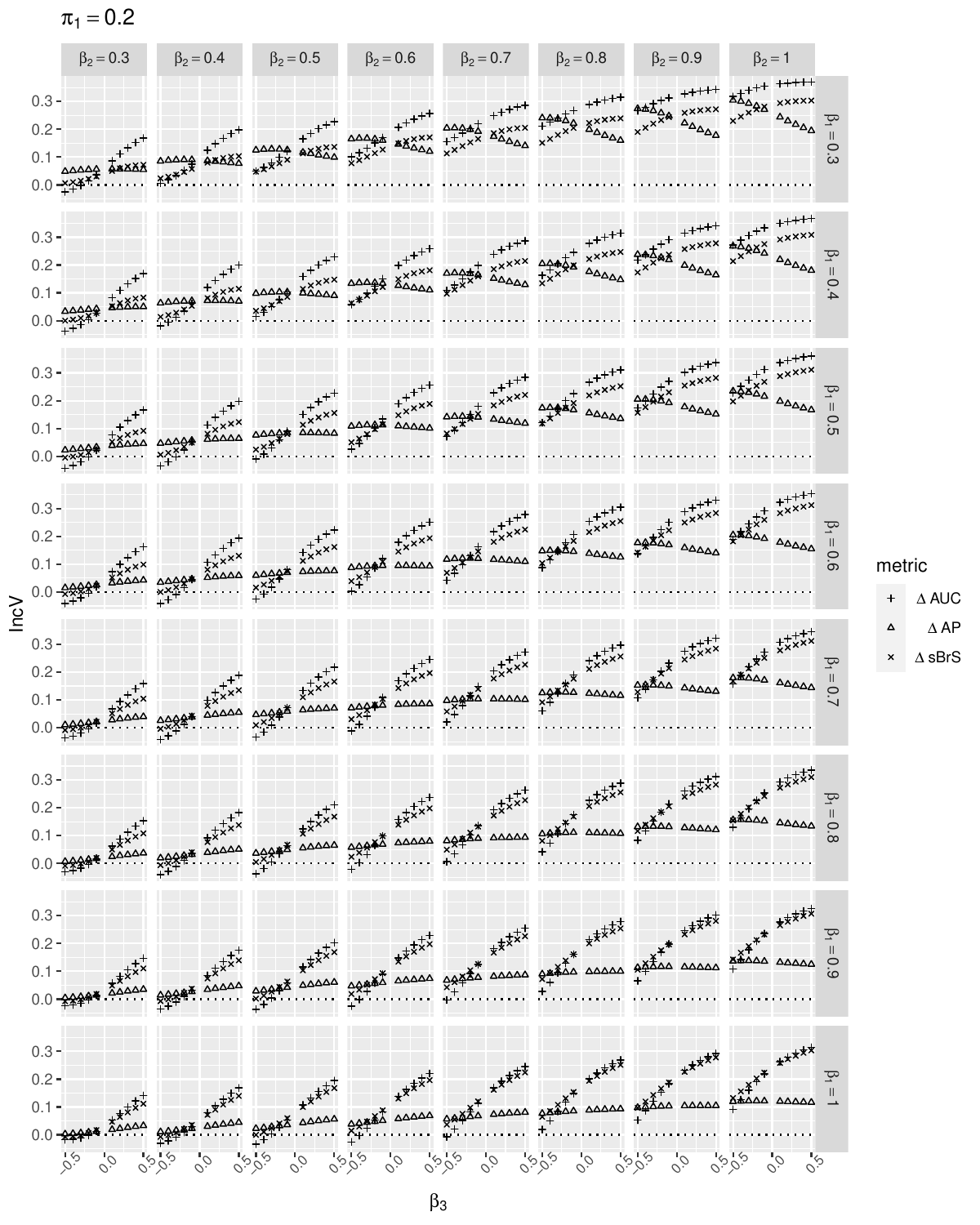}
\caption{The $\Delta \auc$, $\Delta \ap$, and $\Delta \sbs$ under the event rate $\pi=0.2$.}\label{fig:pi-0.2}
\end{figure}

\begin{figure}
\centering
\includegraphics[width=1\textwidth]{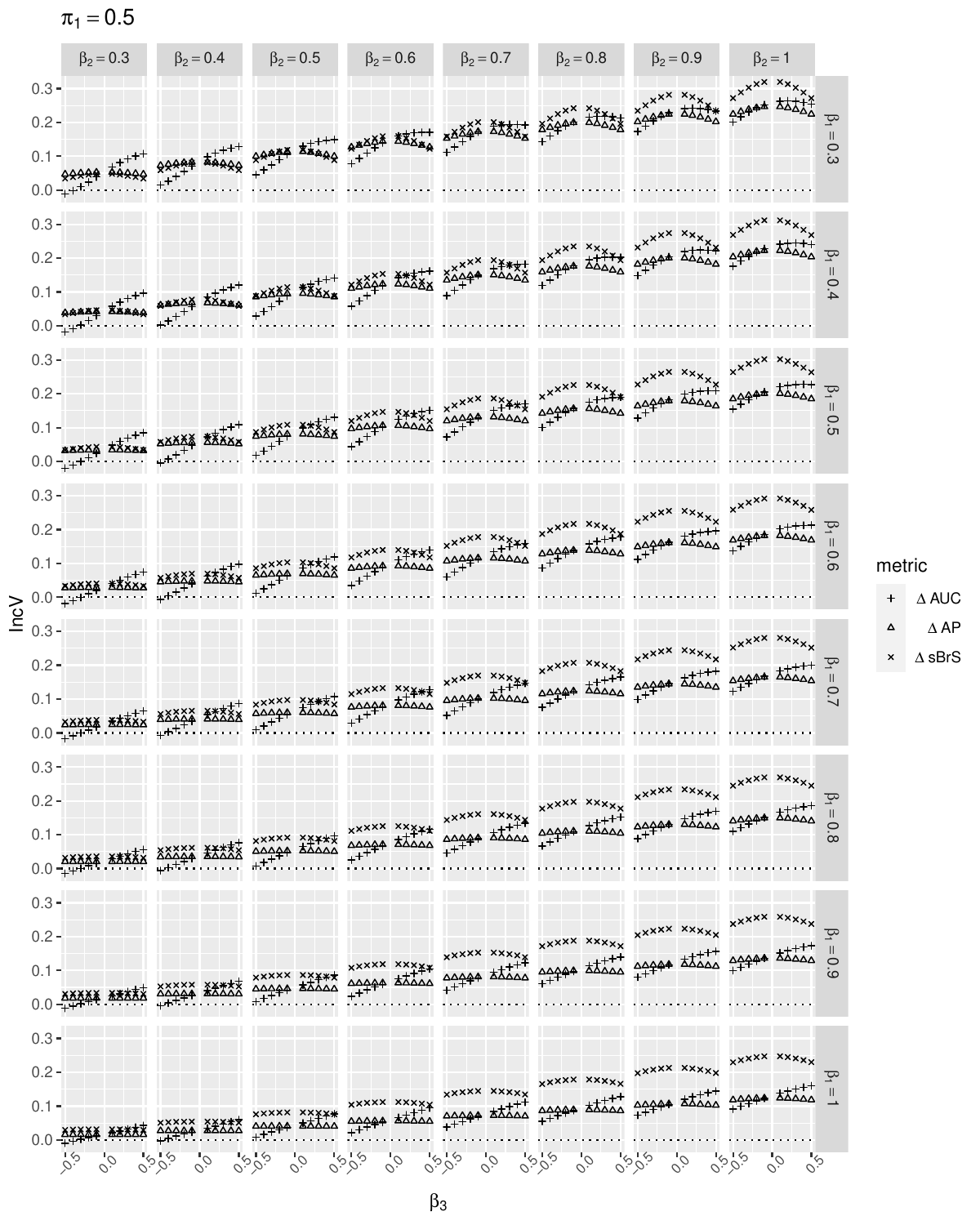}
\caption{The $\Delta \auc$, $\Delta \ap$, and $\Delta \sbs$ under the event rate $\pi=0.5$.}\label{fig:pi-0.5}
\end{figure}

\begin{figure}
\centering
\includegraphics[width=0.9\textwidth]{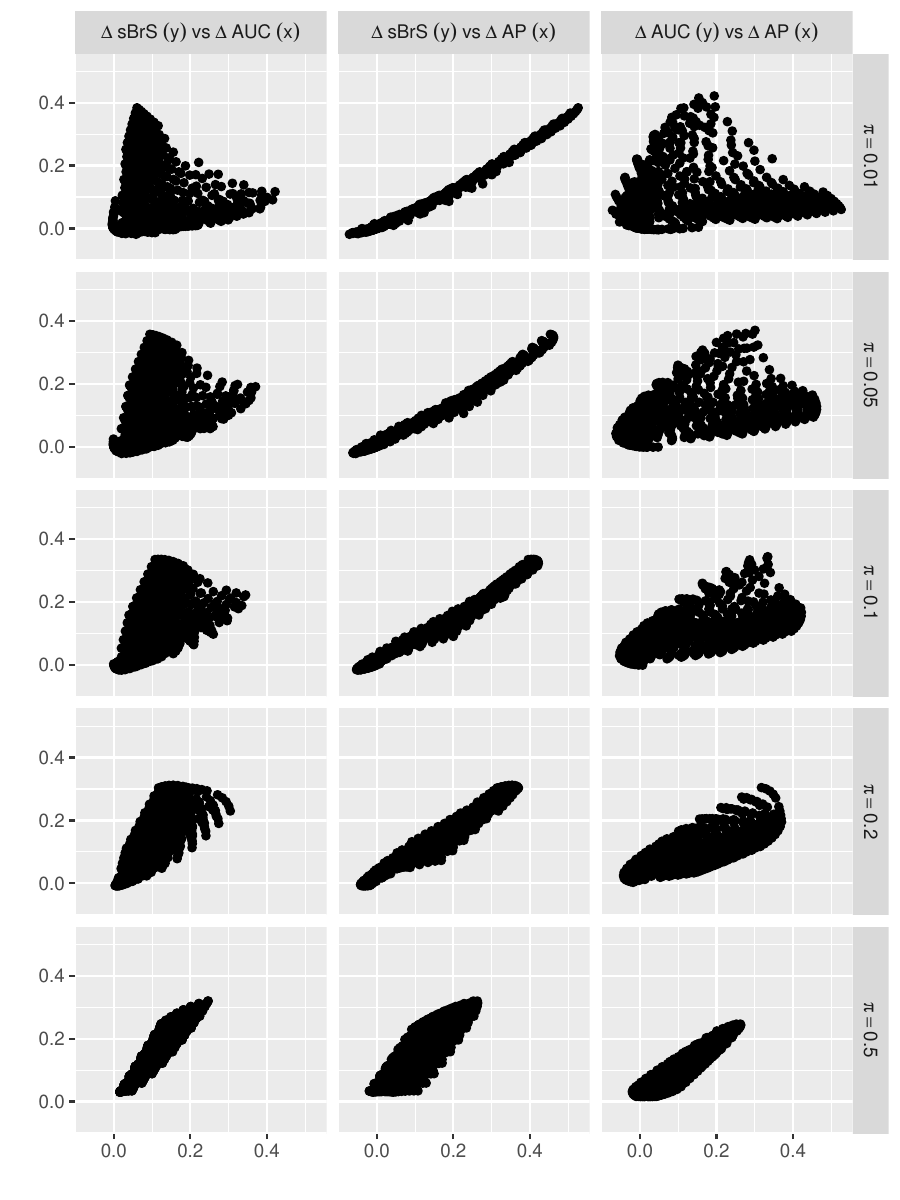}
\caption{Scatter plots of each pair of the IncV metrics under different event rates $\pi$.}\label{fig:NumIll_scatterplot}
\end{figure}

\begin{figure}[!h]
\centering
\includegraphics[width=1\textwidth]{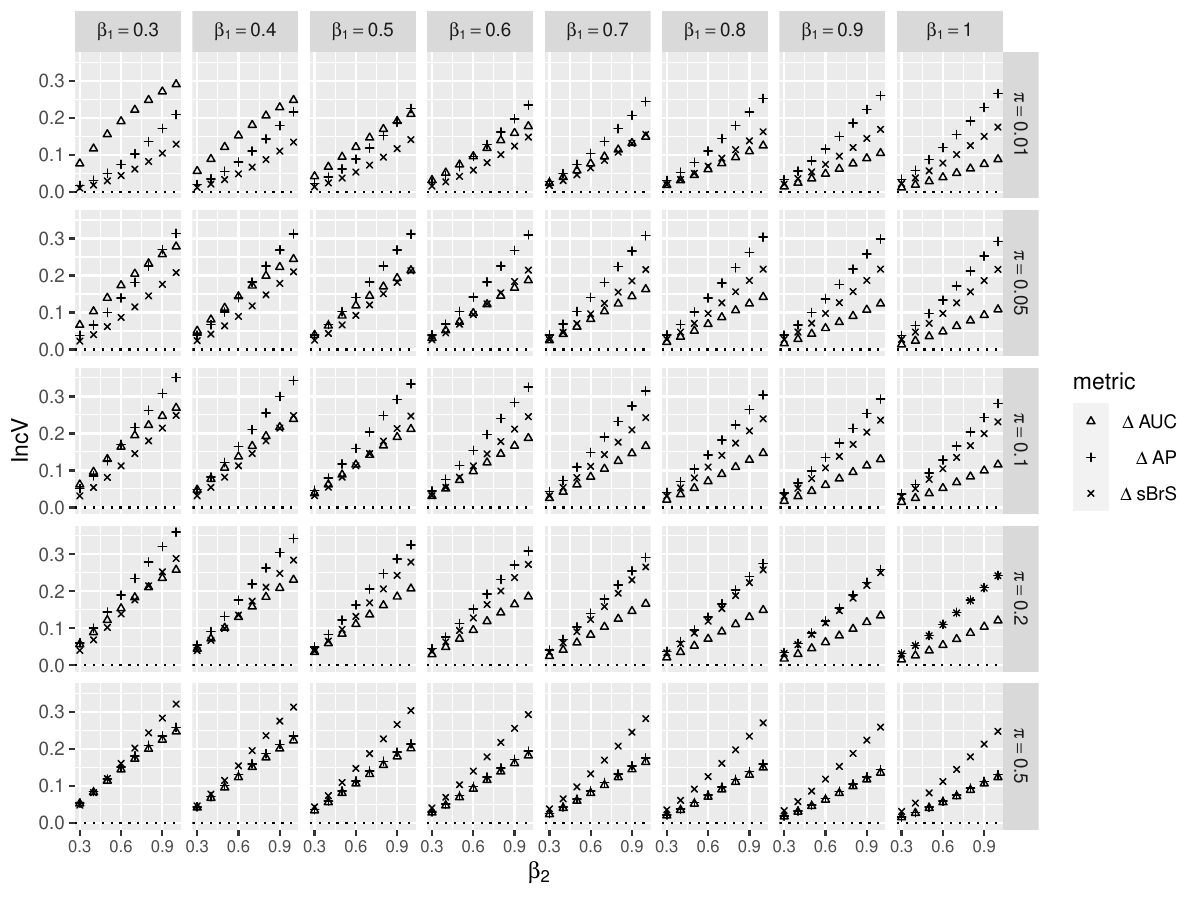}
\caption{Plots of the values of each IncV metric under different event rates $\pi$ when the two-marker model is the true model, i.e., $\beta_3=0$.}\label{fig:NumIll_scatter_true}
\end{figure}

\begin{figure}[!h]
\centering
\includegraphics[width=1\textwidth]{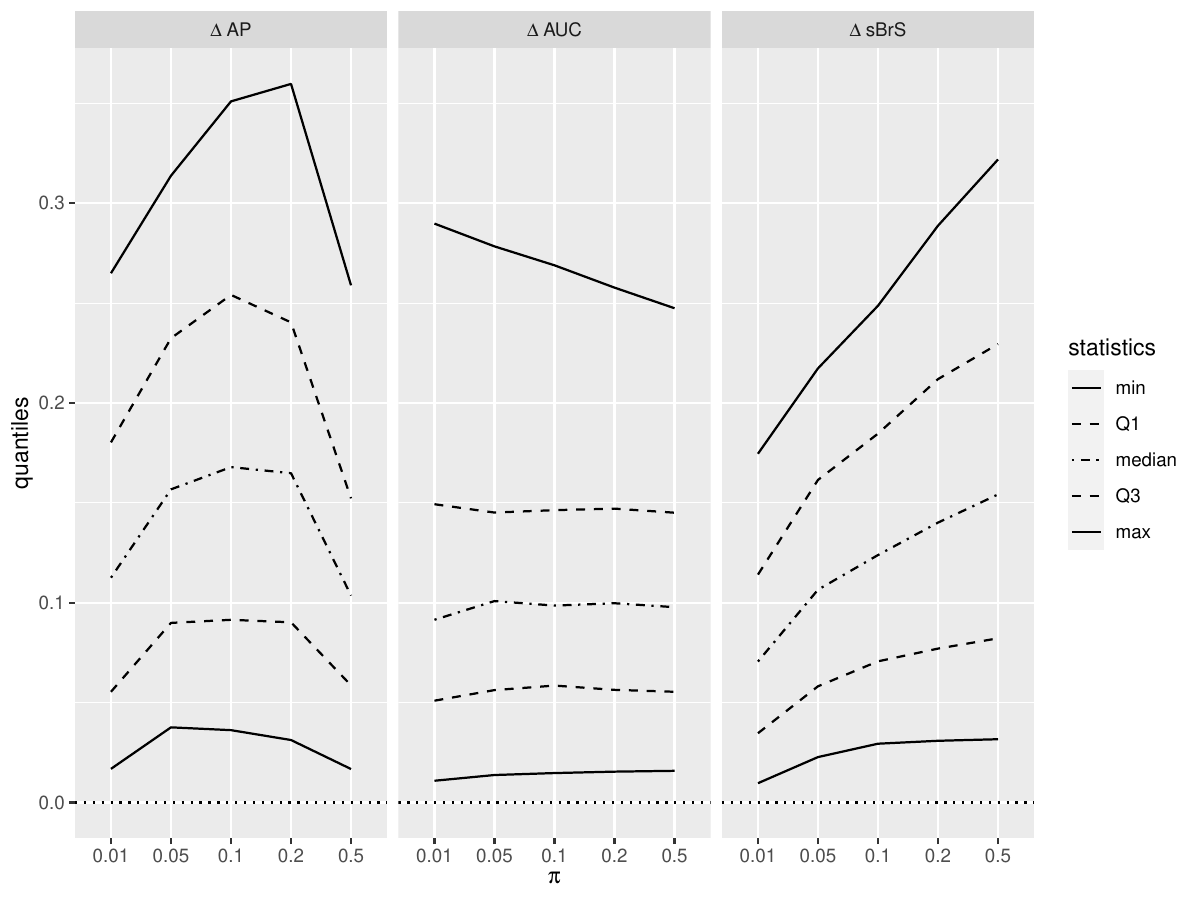}
\caption{Summary statistics of each IncV metric versus different event rates $\pi$ when the two-marker model is the true model, i.e., $\beta_3=0$. The statistics are the minimum, 25-th percentile, median, 75-th percentile, and maximum.}\label{fig:NumIll_quatiles_true}
\end{figure}

In this section, 
\begin{itemize}
\item Figures \ref{fig:pi-0.01} - \ref{fig:NumIll_scatterplot} include the results for the scenarios in which neither of the working risk models is the true model. Figure \ref{fig:pi-0.01} - \ref{fig:pi-0.5} plot the values of the $\Delta\auc$, $\Delta \ap$, and $\Delta \sbs$ for all the scenarios under each of the five event rates. Figure \ref{fig:NumIll_scatterplot} includes the scatter plots of each pair of the IncV metrics by different event rates. 
\item Figure \ref{fig:NumIll_scatter_true}, Figure \ref{fig:NumIll_quatiles_true}, and Table \ref{tab:NumIll_cor} include the results of the scenarios in which the two-marker model is the true model, i.e., $\beta_3=0$, for different event rates. The results include plots of the values of each IncV metric for all the scenarios and different event rates, plots of their summary statistics, and the Pearson correlation of each pair of the IncV metrics.
\end{itemize}

\clearpage
\newpage

\begin{table}[!h]
\centering
\caption{Pearson correlation of each pair of the IncV metrics for different event rates $\pi$ when the two-marker model is the true model, i.e., $\beta_3=0$.}  \label{tab:NumIll_cor}\vspace{0.2cm}
\begin{tabular}{r|r|r|r|r|r}
\hline\hline
Comparison & $\pi=0.01$ &  $\pi=0.05$ & $\pi=0.1$ & $\pi=0.2$ & $\pi=0.5$ \\
\hline
$\Delta \sbs$ vs $\Delta \ap$ & 0.999 & 0.996 & 0.992 & 0.984 & 0.944\\
$\Delta \sbs$ vs $\Delta \auc$ & 0.454 & 0.721 & 0.820 & 0.895 & 0.941\\
$\Delta \auc$ vs $\Delta \ap$ & 0.490 & 0.778 & 0.883 & 0.960 & 1.000\\
\hline
\hline
\end{tabular}
\end{table}

\bibliographystyle{apalike}
\bibliography{AP}